\documentclass[prd,aps,nofootinbib,floatfix,11pt]{revtex4}
\usepackage{amsmath,graphicx,epsfig,amssymb,dsfont,mathtools,}
\usepackage[usenames]{color}
\usepackage{ulem} 
\usepackage{bigstrut}
\usepackage{slashed}
\usepackage{multirow}
\usepackage{subfigure}

\allowdisplaybreaks

\begin{document}\title{The Production of Charm Pentaquark from B meson within SU(3) Analysis}
\author{
   Na Li~$^{1}$~\footnote{Email:TS22180005A31@cumt.edu.cn},Ye Xing~$^{1}$~\footnote{Email:xingye\_guang@cumt.edu.cn},Xiao-hui Hu$^{1, 2}$~\footnote{Email:huxiaohui@cumt.edu.cn}}

\affiliation{$^{1}$  School of Physics, China University of Mining and Technology, Xuzhou 221000, China\\  }
  \affiliation{$^{2}$
  Lanzhou Center for Theoretical Physics and Key Laboratory of Theoretical Physics of Gansu Province, Lanzhou University, Lanzhou 730000, China}
\begin{abstract}
We study the masses and production modes of pentaquark with the quark constituent $c\bar qqqq$, by the tridiquark-diquark model and systematical light flavor quark symmetry SU(3) analysis. The mass splittings show that the S-wave singly charm pentaquark $c\bar udds$, $c\bar dusu$, $c\bar sudu$, and $c\bar sudd$ are below their strong decay thresholds, while $c\bar u ds s, c \bar d uss$ with parity $\frac{1}{2}^-$  are near their strong decay thresholds, which imply the possibility of stable states. Furthermore, we discuss the production of the concerned 15 states from $B$ meson, the analysis within SU(3) symmetry yields the golden channels of production process, $\overline B^0_s\!\!\to   F_{-1/2}^+  \overline p,\ B^-\!\!\to   F_{-3/2}^-  \overline \Lambda^0$, which expected to be work worthily in b-factory.
\end{abstract}
\maketitle

\section{Introduction}
Recently, there is a renewed interest in the study of pentaquark with heavy quark constituents, especially, the LHCb observation in 2015 of hidden charm pentaquark $P_c$ states, $P_c(4380), P_c(4312), P_c(4440)$ and $P_c(4457)$, in the $\Lambda_b^0\to J/\Psi p K^-$~\cite{LHCb:2015yax,LHCb:2019kea}, as well as $P_{cs}(4459)$ and $P_{c}(4337)$ in the $\Xi_b^-\to J/\Psi \Lambda K^-$ and $B_s\to J/\Psi p \bar p$ respectively~\cite{LHCb:2020jpq,LHCb:2021chn}, which have triggered many theoretical studies~\cite{Santopinto:2016pkp,Deng:2016rus,Maiani:2015vwa,Giron:2021fnl,Lebed:2022vks,Azizi:2022qll,Du:2019pij,Wang:2019ato,Chen:2015loa,Eides:2019tgv,Guo:2015umn,Yang:2022bfu,Wang:2022neq,Qin:2022nof,Richard:2019fms}. The existence of pentaquark are proposed by Gell-Mann and Zweig at the birth of quark model in 1964~\cite{Gell-Mann:1964ewy,Zweig:1964ruk}. But until now, only a couple of hidden charm pentaquark come to light. The open heavy flavor ones, for instance open charm tetraquark state $T_{cc}^+$~\cite{LHCb:2021vvq} firstly observed by LHCb in the mass spectrum $D^0D^0\pi^+$ in 2021, therefore are expected similarly. Theoretically, the open flavor pentaquarks with charm quark and $\bar qqqq$ have been discussed by the constituent quark model in Ref.~\cite{Stewart:2004pd}, rough mass estimates are evaluated in the end of paper. The Chromomagnetic Interaction(CMI) model provides more accurate results in Ref.~\cite{An:2020vku}. In addition, the QCD sum rules and chiral effective field theory offer additional methods to investigate the singly heavy pentaquark, individually the mass around 3.21 GeV~\cite{Albuquerque:2013hua,Wang:2018alb,Azizi:2018dva}  and 2.7 GeV~\cite{Wu:2004wg,Yu:2018yxl}  for ground $cudd\bar u$ state.

In the letter, we wound consider the charm pentaquark within the triquark-diquark model. The model is a suitable appropriate~\cite{Karliner:2003sy,Karliner:2003si} to deal with the many quarks system, in this framework, the diquark with spin 0, color anti-triplet, known as good diquark $|(q'q'')^{S(0)}_{C(\bar 3)}>$, and triquark with spin $\frac{1}{2}$, color triplet, builded as $|(cq\bar q)_{C(3)}^{S(\frac{1}{2})}>$. The lowest-lying pentaquark states $\{cq\bar q\}-\{q'q''\}$, whose orbital angular momentum $L=0$ and parity $J^P=\frac{1}{2}^-$, accordingly are divided into two non-singlet color clusters, which combining through the color-triplet binding mechanism~\cite{Lebed:2015tna}.  To investigate the mass splitting of singly charm pentaquark, we adopt the effective Hamiltonian approach, which apart from the constituent quark and diquark masses, including dominant spin-spin and spin-orbit interactions~\cite{Ali:2019clg}.

Note that the light quark symmetry have been successfully adopted in generic hadronic system~\cite{Savage:1989ub,Gronau:1995hm,He:2000ys,Chiang:2004nm,Li:2007bh,Wang:2009azc,Lu:2016ogy,Wang:2017mqp,Wang:2017azm,Shi:2017dto,Wang:2018utj,Huang:2022zsy,Xing:2022aij,Wang:2022kwe,Hu:2022qlr}, which provides a general insight for the decays and productions of hadron, then to be an useful approach to discuss the pentaquark production from B meson. The study is based on the representations of pentaquark, final hadrons and weak transition operator. After constructing the corresponding production Hamiltonian, naturally expanding into transition matrix element, production channels can be attained. A large branching ratios and relatively clear final products wound be better for the choosing golden production channels. We prefer the production channels, which are expected to be observed in b-factory.

The paper is organized as follows. In Sec.II, we present the discussion of the mass splitting of the charm pentaquark. In Sec.III, we analysis the production of the pentaquark states from B meson decays. Some further comments are proposed in Sec.IV.
\section{The mass splitting of the charm pentaquark states}
We study the mass spectrum of S-wave pentaquark states ${c\bar qqqq}(q=u,d,s)$ in the framework of non-relativity doubly heavy triquark-diquark model. Under the diquark-triquark picture, the pentaquark states can be labeled as $|((cq)_{\bar 3}(\bar q)_{\bar 3})_{C= 3}(q'q'')_{C=\bar 3}\rangle$, with two light quarks $q'q''$ into color $\bar 3$ state and singly heavy diquark $cq$ in color $\bar 3$ plus a light anti-quark $\bar q$ in color $\bar 3$. Following with effective Hamiltonian approach~\cite{Ali:2019clg}, the mass spectrum of charm pentaquark with hyperfine structure from spin-spin interactions, can be written as
\begin{eqnarray}\label{eq:hamiltonian}
        \mathcal{H}&=&H_{t}+H_{ld},\nonumber\\
        H_{t}&=&m_{\bar q}+m_{hd}+2 (\mathcal{K}_{cq})_{\bar 3} (S_c\cdot S_q)+2 (\mathcal{K}_{c\bar q})(S_c\cdot S_{\bar q})+2 (\mathcal{K}_{q\bar q})(S_q\cdot S_{\bar q})),\nonumber\\
        H_{ld}&=&m_{ld}+2(\mathcal{K}_{qq'})_{\bar3} (S_{q}\cdot S_{q'})+2(\mathcal{K}_{cq'})_{\bar 3}(S_c\cdot S_q')+2(\mathcal{K}_{qq''})_{\bar 3}(S_q\cdot S_{q''})\nonumber\\&+&2(\mathcal{K}_{cq''})_{\bar 3}(S_c\cdot S_{q''})+2(\mathcal{K}_{qq'})_{\bar 3}(S_q\cdot S_{q'})+2(\mathcal{K}_{q'\bar q})(S_q'\cdot S_{\bar q})+2(\mathcal{K}_{q''\bar q})(S_{q''}\cdot S_{\bar q}).
\end{eqnarray}
The Hamiltonian $H_{t}$ is related with the color triquark, where $m_{\bar q}$ and $m_{hd}$ are the constituent masses of the antiquark and singly charm diquark, respectively. The remanent terms of $H_{t}$ describe the spin-spin interactions in the singly charm diquark and between the diquark constituents and the antiquark. Among the three spin-spin
couplings $(\mathcal{K}_{cq})_{\bar 3} $, $(\mathcal{K}_{c\bar q})$ and $(\mathcal{K}_{q\bar q}) $,  the spin-spin interaction inside the diquark $(\mathcal{K}_{cq})_{\bar 3}$ is argued to be the dominant one.

The $H_{ld}$ term in the Hamiltonian contains the operators responsible for
the spin-spin interaction in the light diquark and its interaction with the triquark. In the singly heavy triquark-diquark system, the suggested spin of singly heavy charm diquark is $S_{cq}=0,1$. Similarily, the spin of ``good" light diquark in the S-wave pentaquark state ${c\bar q qqq}$ is chosen as $S_{q'q''}=0$~\cite{Jaffe:2004ph}. Accordingly, we can write directly the possible configuration of S-wave pentaquark ${cq\bar qqq}$, signed as $|S_{cq},S_t,L_t; S_{q'q''},L_{q'q''};S,L\rangle$.
Here the light diquark $q'q''$ can be any one of the constituents $(ud,du,us,su,ds,sd)$. Except for the spin of singly heavy diquark $S_{cq}$ and light diquark $S_{q'q''}$, the S-wave states with orbital angular momentum $L_t=L_{qq'}=L=0$, the spin of triquark ($(cq) \bar q$) $S_t$ turns out to be ${1}/{2}$ or ${3}/{2}$.

The states with parity $J^P=\frac{1}{2}^-$ and $J^P=\frac{3}{2}^-$, sandwiching the effective mass Hamiltonian Eq.~(\ref{eq:hamiltonian}), then yield the mass spectrum matrix of S-wave pentaquark ${c\bar qqqq}$,
\begin{eqnarray}
&&\mathcal{M}^{L=0}_{J=1/2,1/2,3/2}
=m_{cq}+m_{\bar q}+m_{q'q''}\nonumber\\
&+& \left(\begin{array}{cc} -\frac{3}{2}(\mathcal{K}_{cq})_{\bar3}-\frac{3}{2}(\mathcal{K}_{q'q''})_{\bar3}&\frac{1}{2}\sqrt{3}\mathcal{K}_{c\bar q}+\frac{1}{2}\sqrt{3}\mathcal{K}_{q\bar q}+ \frac{1}{2}\mathcal{K}_{q'\bar q}+\frac{1}{2}\mathcal{K}_{q''\bar q}\\ \frac{1}{2}\sqrt{3}\mathcal{K}_{c\bar q}+\frac{1}{2}\sqrt{3}\mathcal{K}_{q\bar q}+ \frac{1}{2}\mathcal{K}_{q'\bar q}+\frac{1}{2}\mathcal{K}_{q''\bar q}& \frac{1}{2}(\mathcal{K}_{cq})_{\bar3}+\frac{1}{2}(\mathcal{K}_{q'q''})_{\bar3}\end{array}\right).
\end{eqnarray}
In particular, the determination of spin-spin interaction between three spins inside the triquark, i.e., $S_c\cdot S_{\bar q}$ and $S_q\cdot S_{\bar q}$, can be drawn by the wigner 6j-symbols. Further more, for the interaction between triquark and light diquark, such as $S_{\bar q}\cdot S_{q'}$ and $S_{\bar q}\cdot S_{q''}$, it is convenient to utilize the wigner 9j-symbols to describe the recouplings. The remaining step is the numerical calculation. In this work, the spin-spin coupling and the mass of quark and diquark can be taken as in Tab.~\ref{tab:input}.
\begin{table}
    \label{tab:input} \caption{The values of the spin-spin couplings and the masses of quarks and diquarks~\cite{Ali:2019clg,Xing:2021yid}.}
\begin{tabular}{ccccc}\hline\hline
$(\mathcal{K}_{qq'})_{\bar 3}=98$ MeV& $(\mathcal{K}_{ss})_{\bar 3}=23$ MeV
&$(\mathcal{K}_{su/d})_{\bar 3}=59$ MeV& $(\mathcal{K}_{cu/d})_{\bar 3}=15$ MeV&
$(\mathcal{K}_{cs})_{\bar 3}=50$ MeV\\
$\mathcal{K}_{c\bar u/\bar d}=72$ MeV&$\mathcal{K}_{c\bar s}=72$ MeV&$\mathcal{K}_{u\bar d/d\bar u}=318$ MeV
& $\mathcal{K}_{s\bar d/\bar u}=200$ MeV
&$\mathcal{K}_{s\bar s}=103$ MeV\\\hline
 $m_{u/d}=362$ MeV& $m_s=540$ MeV& $m_{ud}=576$ MeV
&$m_{sq}=800$ MeV&$m_{cd/u}=1976$ MeV\\$m_{cs}=2105$ MeV&$m_{ss}=1099$ MeV \\\hline\hline
    \end{tabular}
\end{table}
Certainly, one should consider the uncertainties from these couplings and masses, we take $10\%$ as the error in this work.

We diagonalize the mass matrix and obtain the mass splittings of pentaquark  ${c\bar qqqq}$ shown in Tab.~\ref{tab:mass}. The results from Chromomagnetic Interaction(CMI) model, QCD sum rules(QCDSR) and chiral effective field theory(ChEFT) are collected for comparison. Our study is similar with the calculation from CMI, but is different from the results of QCDSR and ChEFT, especially for the ground states with parity $\frac{1}{2}^-$. In our considering, both components $c\bar s snn,~c\bar nssn$ and $c\bar nsnn$ are lower than their strong thresholds $\Xi_{c}K,~\Omega_{c}\pi$ and $\Xi_{c}\pi$ respectively, further more, the mass of the component $c\bar sssn$ is close to the strong thresholds $\Xi_{c}\eta$. It is interesting that the pentaquark with constituent $c\bar u ds s, c \bar d uss$ with parity $\frac{1}{2}^-$  are near their strong decay thresholds $\Omega_c \pi$, which may imply the possibility of stable states. The pentaquark with constituent $c\bar u ds s, c \bar d uss$ near the corresponding threshold $\Omega_c \pi$, respectively below that about $178$ MeV and $178$ MeV, which wound be expected to be verified in future experiments. Especially, the components $c\bar snnn$, $c\bar nnnn$ in our work are lower than the strong threshold $\Lambda_{c} K$, $\Lambda_{c}\pi$ about 319 MeV, 416 MeV, while CMI show a different result, which is near that about several MeV.
\begin{table}
    \centering
    \caption{The mass splitting of the S-wave pentaquark $P_{c\bar qqqq}$ coming from the hyperfine structures of triquark $|((cq)_{\bar 3}^{0}\bar q_{\bar 3}^{1/2})_{C=3}^{S=1/2}\rangle$, $|((cq)_{\bar 3}^{1}\bar q_{\bar 3}^{1/2})_{C=3}^{S=1/2}\rangle$, and $|((cq)_{\bar 3}^{0}\bar q_{\bar 3}^{1/2})_{C=3}^{S=3/2}\rangle$,  whose total parities are $J^P=\frac{1}{2}^-, \frac{1}{2}^-$ and $\frac{3}{2}^-$ respectively, where the superscript represents spin of diquark or triquark and subscript shows color multiplet. $n$ represents the light quark $u,d$. $I$ is the isospin of pentaquark. The work are  compared with the Chromomagnetic Interaction(CMI) model, QCD sum rules(QCDSR)~\cite{Albuquerque:2013hua,Wang:2018alb,Azizi:2018dva} and chiral effective field theory(ChEFT), the corresponding strong thresholds are listed in the last column.}\label{tab:mass}
       \begin{tabular}{|l|c|c|c|c|c|c}\hline
      \multirow{2}*{Pentaquark($J^P,I$)} &  \multicolumn{4}{c|}{Mass(GeV)} 	& \multirow{2}*{threshold}  \\ \cline{2-5}
      &  this work  & CMI~\cite{An:2020vku}  & QCDSR & ChEFT &     \\\hline
    $c\bar n  nnn (\frac{1}{2}^-,0/1/2)$ &  $2.005\pm{0.182}$ & 2.446 & $-$&2.704\cite{Wu:2004wg}& $ \Lambda_{c}\pi$ \\\hline
    $c\bar s snn (\frac{1}{2}^-,0/1/2)$ &  $2.733\pm{0.151}$ &  2.781 & $-$&$-$&$\Xi_{c}K$ \\\hline
    $c\bar nsnn (\frac{1}{2}^-,\frac{1}{2}/\frac{3}{2})$ &  $2.300\pm{0.194}$ &  2.404 &$-$&2.791\cite{Yu:2018yxl}&$\Xi_{c}\pi$  \\\hline
    $c\bar s ssn (\frac{1}{2}^-,\frac{1}{2}/\frac{3}{2})$ &  $3.015\pm{0.145}$ & 3.103 & $-$&$-$&$\Xi_{c}\eta$ \\\hline
  $c\bar nssn (\frac{1}{2}^-,0/1)$ &  $2.657\pm{0.135}$ & 2.776 &$-$&3.050\cite{Huang:2018wgr}& $\Omega_{c}\pi$ \\\hline
 $c\bar snnn (\frac{1}{2}^-,\frac{1}{2}/\frac{3}{2})$ &  $2.465\pm{0.187}$ & 2.779 &$-$&$-$& $\Lambda_{c}K$ \\\hline
    \hline
    $c\bar nn nn (\frac{1}{2}^-,0/1/2)$ &  $3.313\pm{0.192}$ & 2.631 & $-$&$-$& $ \Lambda_{c}\pi$ \\\hline
    $c\bar ss nn (\frac{1}{2}^-,0/1/2)$ &  $3.539\pm{0.163}$ &  2.930 &$-$&$-$& $\Xi_{c}K$ \\\hline
    $c \bar nsnn (\frac{1}{2}^-,\frac{1}{2}/\frac{3}{2})$ &  $3.444\pm{0.197}$ &  2.575 &$-$&2.937\cite{Yu:2018yxl}&$\Xi_{c}\pi$ \\\hline
    $c\bar ss sn (\frac{1}{2}^-,\frac{1}{2}/\frac{3}{2})$ &  $3.595\pm{0.155}$ & 3.272 & $-$&$-$&$\Xi_{c}\eta$ \\\hline
  $c\bar nssn (\frac{1}{2}^-,0/1)$ &  $3.747\pm{0.143}$ &3.114 &$-$&3.119\cite{Huang:2018wgr}& $\Omega_{c}\pi$ \\\hline
  $c\bar snnn (\frac{1}{2}^-,\frac{1}{2}/\frac{3}{2})$ &  $3.336\pm{0.198}$ & 2.908 &$-$&$-$& $\Lambda_{c}K$ \\\hline
   $c\bar nn nn (\frac{3}{2}^-,0/1/2)$ &  $3.103\pm{0.185}$ &  2.540 &$-$&2.802\cite{Yu:2018yxl}&  $ \Lambda_{c}^*\pi$ \\\hline
    $c\bar ss nn (\frac{3}{2}^-,0/1/2)$ &  $3.468\pm{0.175}$ &  3.043&$-$&$-$&$\Xi_{c}^*K$ \\\hline
    $c \bar nsnn (\frac{3}{2}^-,\frac{1}{2}/\frac{3}{2})$ &  $3.174\pm{0.198}$ &  2.666 &$-$&2.912\cite{Yu:2018yxl}& $\Xi_{c}^*\pi$ \\\hline
    $c\bar s ssn (\frac{3}{2}^-,\frac{1}{2}/\frac{3}{2})$ &  $3.563\pm{0.153}$ & 3.208&$-$&$-$ & $\Xi_{c}^*\eta$ \\\hline
  $c\bar nssn (\frac{3}{2}^-,0/1)$ &  $3.637\pm{0.144}$ & 2.865  &3.2&$-$& $\Omega_{c}^*\pi$\\\hline

    \end{tabular}
  \end{table}
\begin{figure}
  \centering
  \includegraphics[width=0.88\columnwidth]{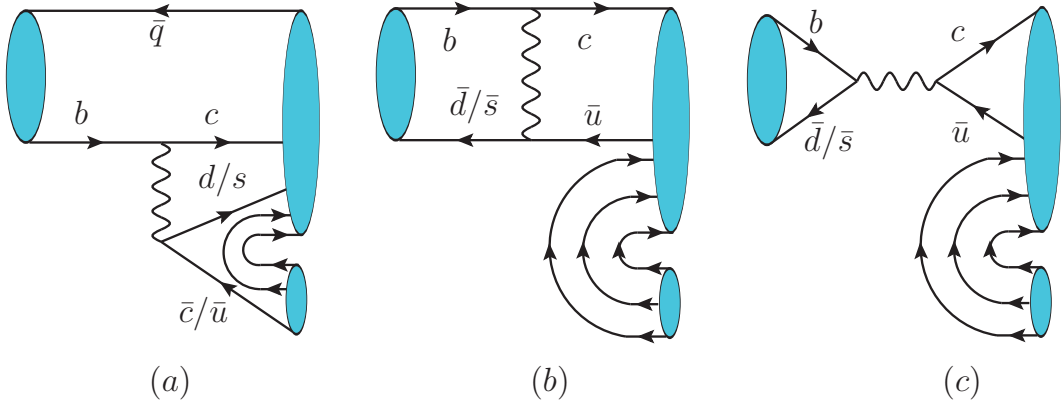}\\
  \caption{The typical topology diagrams for the production of charm pentaquark $P_{c\bar qqqq}$ from the bottom meson. The production depends on the weak decay of b quark, which leads to different topologies (a,b,c) including charm pentaquark, and light anti-baryon or anti-charm anti-baryon
    in final states.}\label{fig:production}
\end{figure}
\section{Productions from $B$ Mesons}\label{sec:production}

\subsection{Representations of Charm Pentaquark}
Light quarks satisfy the SU(3) flavor symmetry, and behave well at the level of hadrons. We can use group representations to describe the hadrons, in consideration of individual spin or orbital quantum number. We can transform the singly charm pentaquark $c \bar q qqq$ under the SU(3) symmetry,
\begin{eqnarray}
3 \otimes 3 \otimes 3 \otimes \bar{3}=\bar{3}\oplus \bar{3} \oplus 6 \oplus 15 \oplus{6} \oplus \bar{3} \oplus 6 \oplus 15 \oplus 24 .
\end{eqnarray}
After group decomposition above, we get the irreducible representations of new combination states $6$, $15$ and $24$. By tensor reduction, the irreducible representations can be expressed as different tensor forms that labeled with $T_i^{jkl}$,
\begin{eqnarray}\label{eq:tensor}
T_i^{jkl} &=&(\overline{T}_{15})_{\{i \alpha\}}^{l} (A_1)^{\alpha jk}
+(\overline{T}_{15})_{\{i \alpha\}}^{\beta} (A_2)_{i \alpha \beta \gamma }^{i k \alpha  \gamma l}
+(\widetilde{T}_{6})^{\{l \beta\}} (B_1)_{i \alpha \beta}^{ \alpha jk}
+(\widetilde{T}_{6})^{\{mn\}} (B_2)_{i \alpha \beta m n}^{ \alpha \beta ijk}\nonumber\\
&&+(\hat{T}_{ \bar{3}})_{m} (C_1)_{i \alpha \beta }^{ \alpha \beta jklm}
+(\hat{T}_{ \bar{3}})_{\alpha} (C_2)_{i \gamma }^{ \alpha \gamma jkl}
+(\hat{T}_{ \bar{3}})_{\alpha} (C_3)_{i }^{ \alpha  jkl}
+({T}_{24})_{i}^{\{ jkl \} } (D_1),
\end{eqnarray}
and the coefficients of irreducible representations can be taken as follows,
\begin{eqnarray*}
&&(A_1)^{\alpha jk}=\frac{1}{2}\varepsilon^{\alpha jk},\
(A_2)_{i \alpha \beta \gamma }^{i k \alpha  \gamma l}=\frac{1}{3}(\delta_{\beta}^i \delta_{\gamma}^k+\delta_{\beta}^k \delta_{\gamma}^i) \varepsilon^{\alpha \gamma l},\\
&&(B_1)_{i \alpha \beta}^{ \alpha jk}=\frac{1}{4} \varepsilon^{\alpha jk} \varepsilon_{i \alpha \beta},\
(B_2)_{i \alpha \beta m n}^{ \alpha \beta ijk}=\frac{1}{6}\varepsilon_{i \alpha n}\varepsilon^{\alpha \beta l (\delta_{\beta}^k \delta_{m}^i+\delta_{\beta}^j \delta_{m}^j)},\\
&&(C_1)_{i \alpha \beta }^{ \alpha \beta jklm}=\frac{1}{8}\varepsilon^{\alpha jk} \varepsilon_{i\alpha \beta} \varepsilon^{ml \beta},\
(C_2)_{i \gamma }^{ \alpha \gamma jkl}=-\frac{1}{18}(\delta_{i}^j \delta_{r}^k+\delta_i^k \delta_{\gamma}^j) \varepsilon^{\alpha \gamma l},\\
&&(C_3)_{i }^{ \alpha  jkl}=\frac{1}{36}\Big( 6\delta_i^{\alpha} \varepsilon^{jkl}- \delta_{i}^j\varepsilon^{\alpha kl}-\delta_{i}^k\varepsilon^{\alpha jl} \Big),\
(D_1)=\frac{1}{3}.
\end{eqnarray*}
The new combination states $6$, $15$ and $24$ can be expressed as irreducible representations ${T}_{6}$, ${T}_{15}$ and $T_{24}$. Here the anti-symmetry index can be identified as $[ij]$, and the symmetry indexes can be signed with $\{ij\}$. The coefficients consist of tensor $\delta$ and antisymmetric tensor $\varepsilon$.
We can get the quark components of the states in flavor space by expanding the tensor representations which have been deduced above, and the nonzero components are listed in Tab~\ref{tab:bar6}. We also give the weight graphs of states $\bar 6$,  $15$ and $24$ in Fig~\ref{fig:weight1}, whose flavor structures given in Appendix.\ref{sec:tensorform}. The six states $F_{-3/2}^{-}$, $F_{3/2}^{++}$, $F_{1/2}^{++}$, $F_{-1/2}^{+}$, $ F_{-1}^-$ and $F_{1}^+$ in the 15 state, do not contain ${q\bar q}$ pair, so these six states are relatively stable~\cite{Stewart:2004pd} and wound be the concerned states in the work.
\begin{eqnarray*}
F_{-3/2}^{-} = c\bar{u}\text{dsd}-c\bar{u}\text{sdd},\quad
F_{-1}^- = c\bar{u}\text{dss}-c\bar{u}\text{sds},\quad
F_{3/2}^{++} = c\bar{d}\text{suu}-c\bar{d}\text{usu} ,\\
F_{1}^+ = c\bar{d}\text{sus}-c\bar{d}\text{uss} ,\quad
F_{1/2}^{++} = c\bar{s}\text{udu}-c\bar{s}\text{duu} ,\quad
F_{-1/2}^{+} = c\bar{s}\text{udd}-c\bar{s}\text{dud} .
\end{eqnarray*}
Immediately, the masses of $F_{-3/2}^{-}$, $F_{3/2}^{++}$, $F_{1/2}^{++}$ and $F_{-1/2}^{+}$ are found respectively as $2.300$ GeV, $2.300$ GeV, $2.465$ GeV and $2.465$ GeV. Obviously they are lower than their corresponding strong thresholds $\Xi_{c}^0\pi^-$, $\Xi_{c}^+\pi^+$, $\Lambda_{c}^+K^+$ and $\Lambda_{c}^+K^0$ with $308$ MeV, $308$ MeV, $319$ MeV and $319$ MeV. While the masses of $F_{-1}^-$ and $F_{1}^+$ are both $2.835$ GeV, nearing their strong thresholds $\Omega_{c}^{0}\pi^-$ about $178$ MeV, indicating that they may be stable pentaquark states. However the sizeable value of the inaccurate coming from the spin-spin coupling $\mathcal{K}$ coefficients and the mass of diquark $m_{cq}$, are large enough to change the decay behaviour of the states. Accordingly the stability of the states $F_{-1}^-$ and $F_{1}^+$ remains to be an open question. Nevertheless, the $F_{-3/2}^-,~ F_{3/2}^{++},~F_{1/2}^{++}$ and $F_{-1/2}^{+}$ with less controversial in our work, should be weakly decay, which expected to recognized in future experiments.
\begin{figure}
\includegraphics[scale=1,width=1\textwidth]{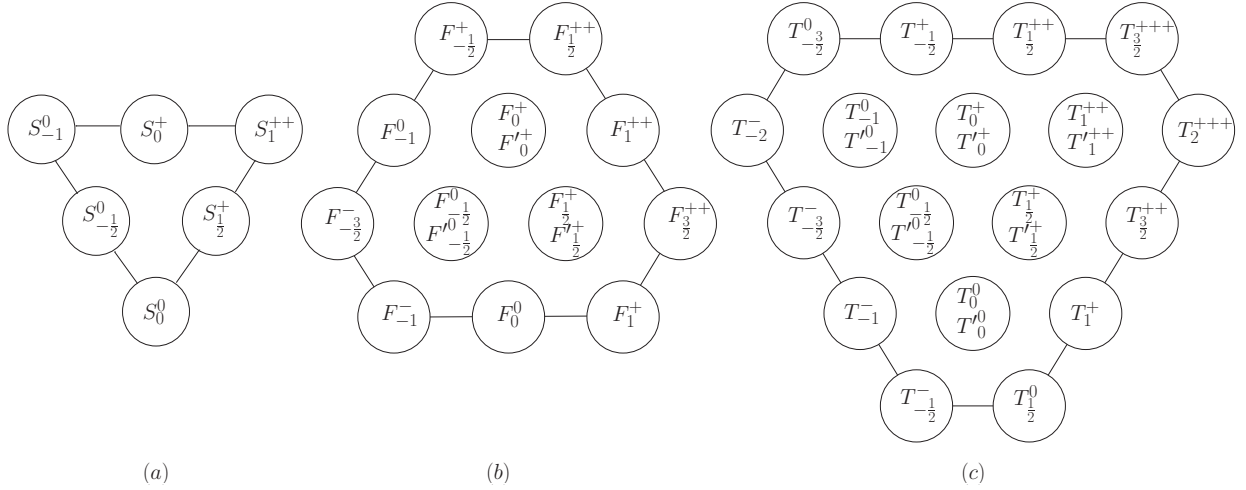}
\caption{The weight diagrams(a-c) show the $6$, $15$, $24$ multiple states of singly charm pentaquark, signed as $S_6, F_{15}, T_{24}$.}\label{fig:weight1}
\end{figure}
\begin{table}
\caption{The possible representations of  pentaquark $c \bar q qqq$ with states $ 6 $, $15$ and $ 24 $. Let $S$, $F$ and $T$ be the names of the states respectively. Meanwhile, we give the tensor representations $T_{6}/T_{15}/T_{24}$, isospin $I_3$ and hyper-charge Y of the corresponding states.}\label{tab:bar6}
\begin{tabular}{|c |c c c c|c c c c| c c c c|}\hline\hline
States& Name & Tensor& $I_3$ & $Y$&Name & Tensor& $I_3$ & $Y$&Name & Tensor& $I_3$ & $Y$ \\\hline
\multirow{2}{*}{6 state}&
$S_{0}^{0}$& $T_{\{33\}}$ & 0&0&
$S_{-\frac{1}{2}}^{0}$& $T_{\{23\}}$ & $-\frac{1}{2}$&1&
$S_{\frac{1}{2}}^{+}$& $T_{\{13\}}$ & $\frac{1}{2}$&1\\
&$S_{-1}^{0}$& $T_{\{22\}}$ & -1&2&
$S_{0}^{+}$& $T_{\{12\}}$ & 0& 2 &
$S_{1}^{++}$& $T_{\{11\}}$ & 1&2
\\\hline
\multirow{4}{*}{15 state}&
$F_{-1}^{-}$& $T_3^{\{11\}}$ & -1&0&
$F_{0}^{0}$& $T_3^{\{12\}}$ & 0&0&
$F_{1}^{+}$& $T_3^{\{22\}}$ & 1&0\\&
$F_{-\frac{3}{2}}^{-}$& $T_2^{\{11\}}$ & $-\frac{3}{2}$&1&
$F_{-\frac{1}{2}}^{0},{F'}_{-\frac{1}{2}}^{0}$& $T_2^{\{12\}},T_3^{\{13\}}$ & $-\frac{1}{2}$&1&
$F_{\frac{1}{2}}^{+},{F'}_{\frac{1}{2}}^{+}$& $T_1^{\{12\}},T_3^{\{23\}}$ & $\frac{1}{2}$&1\\&$F_{\frac{3}{2}}^{++}$& $T_1^{\{22\}}$ & $\frac{3}{2}$&1&
$F_{-1}^{0}$& $T_2^{\{13\}}$ & $-1$&2&
$F_{0}^{+},{F'}_{0}^{+}$& $T_1^{\{13\}},T_2^{\{23\}}$ & $0$&2\\&
$F_{1}^{++}$& $T_1^{\{23\}}$ & $1$&2&
$F_{-\frac{1}{2}}^{+}$& $T_2^{\{33\}}$ & $-\frac{1}{2}$&3&
$F_{\frac{1}{2}}^{++}$& $T_1^{\{33\}}$ & $\frac{1}{2}$&3
\\\hline
\multirow{6}{*}{24 state}
&$T_{-\frac{3}{2}}^{0}$& $T_3^{\{222\}}$ & $-\frac{3}{2}$&3
&$T_{-\frac{1}{2}}^{+}$& $T_3^{\{122\}}$ & $-\frac{1}{2}$&3
&$T_{\frac{1}{2}}^{++}$& $T_3^{\{112\}}$ & $\frac{1}{2}$&3\\
&$T_{\frac{3}{2}}^{+++}$& $T_3^{\{111\}}$ & $\frac{3}{2}$&3

&$T_{-2}^{-}$& $T_1^{\{112\}}$ & -2&2
&$T_{-1}^{0},{T'}_{-1}^{0}$& $T_1^{\{122\}},T_3^{\{223\}}$ & -1&2\\
&$T_{0}^{+},{T'}_{0}^{+}$& $T_1^{\{112\}},T_2^{\{112\}}$ & 0&2
&$T_{1}^{++},{T'}_{1}^{++}$& $T_2^{\{112\}},T_3^{\{113\}}$ & 1&2
&$T_{2}^{+++}$& $T_2^{\{111\}}$ & 2&2\\

&$T_{-\frac{3}{2}}^{-}$& $T_1^{\{223\}}$ & $-\frac{3}{2}$&1
&$T_{-\frac{1}{2}}^{0},{T'}_{-\frac{1}{2}}^{0}$& $T_2^{\{223\}},T_3^{\{233\}}$ & $-\frac{1}{2}$&1
&$T_{\frac{1}{2}}^{+},{T'}_{\frac{1}{2}}^{+}$& $T_1^{\{113\}},T_3^{\{133\}}$ & $\frac{1}{2}$&1\\
&$T_{\frac{3}{2}}^{++}$& $T_2^{\{113\}}$ &$\frac{3}{2}$&1

&$T_{-1}^{-}$& $T_1^{\{233\}}$ & -1&0
&$T_{0}^{0},{T'}_{0}^{0}$& $T_1^{\{133\}},T_2^{\{223\}}$ & $0$&0\\
&$T_{1}^{+}$& $T_2^{\{133\}}$ & 1&0

&$T_{-\frac{1}{2}}^{-}$& $T_1^{\{333\}}$ & $-\frac{1}{2}$&-1
&$T_{\frac{1}{2}}^{0}$& $T_2^{\{333\}}$ & $\frac{1}{2}$&-1
\\\hline
\end{tabular}
\end{table}

\subsection{The Production by SU(3) analysis}
According to the analysis of SU(3) light quark flavor symmetry, we will discuss the possible production and decay modes of the pentaquark states $c\bar qqqq$ in this subsection. This production can be realized by studying the weak decay of B meson. At the stage of decay modes, we focus on the explores of stable pentaquark candidates, which give priority to the weak decays similarly.
The weak interaction of production for the pentaquark states, induced by the transition $b\to c\bar qq/c\bar cq$, can be classified  by the quantities of CKM matrix elements.
For non-leptonic decays of $b$ quark, we classify the transitions into two groups.
\begin{eqnarray}
b\to  c\bar cs/c\bar ud , \;\;\;
b\to  c \bar c d/c\bar u  s , \nonumber
\end{eqnarray}
which are Cabibbo allowed, and singly Cabibbo suppressed transitions respectively.
The transition $b \to  c  \bar q q$  can be decomposed as ${\bf  3}\otimes {\bf\bar3}={\bf  1}\oplus {\bf  8}$. The transition $b \to  c  \bar c  q$  can be decomposed as ${\bf  3}$.
Here we offer the nonzero SU(3) tensor components of Cabibbo allowed transition given as  consistently. In the quark level, there are two production processes of the pentaquark states from B meson decay. The operator of the transition  $b\to  c\bar cd/s $ form an triplet, with $(H_{3})^2=V_{cd}^*,~(H_{3})^3=V_{cs}^*$. And the operator of the transition $b\to  c \bar u  d/s $ can form an octet 8, whose nonzero composition followed as $(H_{8})_{1}^2=V_{ud}^*,~(H_{8})_{1}^3=V_{us}^*$.

Among the calculation of the production with SU(3) symmetry analysis, the representations of initial and final states are essential inputs. The initial states, B meson, including a bottom quark and one light quark, $\overline B^i=\big(\begin{array}{ccc}B^-, &\overline B^0, &\overline B^0_s  \end{array} \big)$. The representations of light anti-baryons can be decomposed into an octet and an anti-decuplet~\cite{Wang:2017azm}. The decuplet can be referred to~\cite{Xing:2021yid}, the octet can be written as
\begin{eqnarray}
F_8= \left(\begin{array}{ccc} \frac{1}{\sqrt{2}}\overline\Sigma^0+\frac{1}{\sqrt{6}} \overline\Lambda^0 & \overline\Sigma^- & \overline \Xi^-   \\ \overline\Sigma^+  &  -\frac{1}{\sqrt{2}} \overline\Sigma^0+\frac{1}{\sqrt{6}}\overline\Lambda^0 & \overline\Xi^0 \\ \overline p   & \overline n  & -\sqrt{\frac{2}{3}}\overline\Lambda^0
  \end{array} \right).
\end{eqnarray}
The representation of singly anti-charm anti-baryons can be decomposed into one triple states and anther anti-sextet, which can be written respectively as
\begin{eqnarray}
F_{c3}= \left(\begin{array}{ccc} 0 & \overline \Lambda_{\bar c}^-  &  \overline \Xi_{\bar c}^-   \\ -\overline \Lambda_{\bar c}^- &  0 & \overline \Xi_{\bar c}^0\\ \ -\overline \Xi_{\bar c}^-  & -\overline \Xi_{\bar c}^0& 0
  \end{array} \right),\quad\quad
F_{c\bar6}= \left(\begin{array}{ccc}\overline \Sigma_{\bar c}^{--} & \frac{1}{\sqrt{2}}\overline \Sigma_{\bar c}^{-} & \frac{1}{\sqrt{2}} \overline \Xi_{\bar c}^{\prime-}   \\ \frac{1}{\sqrt{2}}\overline \Sigma_{\bar c}^- &  \overline \Sigma_{\bar c}^0 & \frac{1}{\sqrt{2}} \overline \Xi_{\bar c}^{\prime0}\\ \ \ \frac{1}{\sqrt{2}}\overline \Xi_{\bar c}^{\prime-}  &  \frac{1}{\sqrt{2}}\overline \Xi_{\bar c}^{\prime 0}& \overline \Omega_{\bar c}^{0}
  \end{array} \right).
\end{eqnarray}
The possible Hamiltonian for the production of concerned pentaquark ground $15$ states from one $B$ meson, induced by the transition $b\to c \bar{c}d/s$ and $b\to c \bar{u}d/s$ in the quark level, can be writen directly as,
\begin{eqnarray}
 {\mathcal{H}}_{15} &=& a_{1}{B}_{i}(H_{3})^{j}({\overline T}_{15})_{j}^{\{ik\}}(\overline F_{c3})_{k}+b_{1}{B}_{i}(H_{3})^{k}({\overline T}_{15})_{j}^{\{i\alpha\}}(\overline F_{c\bar 6})^{\{jl\}}\ \varepsilon_{\alpha kl}+c_{1}{B}_{i}(H_{8})_{j}^{i}(\overline T_{15})_{k}^{\{jl\}}(\overline F_{8})_{l}^{k}\ \nonumber\\
 &+& c_{2}{B}_{i}(H_{8})_{j}^{k}(\overline T_{15})_{k}^{\{il\}}(\overline F_{8})_{l}^{j}+c_{3}{B}_{i}(H_{8})_{j}^{k}(\overline T_{15})_{k}^{\{jl\}}(\overline F_{8})_{l}^{i}+c_{4}{B}_{i}(H_{8})_{j}^{l}(\overline T_{15})_{k}^{\{ij\}}(\overline F_{8})_{l}^{k}\ \nonumber\\ &+& d_{1}{B}_{j}(H_{8})_{k}^{i}({\overline T}_{15})_{l}^{\{j\alpha\}}(\overline F_{10})^{klm}\ \varepsilon_{\alpha im}+d_{2}{B}_{k}(H_{8})_{j}^{i}({\overline T}_{15})_{l}^{\{j\alpha\}}(\overline F_{10})^{klm}\ \varepsilon_{\alpha im}.\label{eq:hamiltonian6}
 \end{eqnarray}
The parameters $a_{i}$ ($b_{i}$ or $c_{i}$) with $i=1,2,3$, are the non-perturbative coefficients. $F_{c3}$ and $F_{c\bar 6}$ are the triplet and anti-sextet of anti-charm anti-baryon, while $F_{8}$ and $F_{\overline {10}}$ are octet and decuplet of the light anti-baryon. The concerned pentaquark ground states are noted with ${T}_{15}$, $B$ represents the bottom meson. In the quark level, the productions of singly charm pentaquark multiple states can be described with topological diagrams, as shown with Fig.~\ref{fig:production}.
We expand the Hamiltonian and collect the possible processes of concerned pentaquark 15 states, gathering into Tab.~\ref{tab:F15concern}, as well as the fully results shown in Tab.\ref{tab:P15full1} and Tab.\ref{tab:P15full2} of Appendix.\ref{sec:tables}. Meanwhile, it is ready to reduce the relations of decay widths between different channels, which given as follows.
\begin{align*}
&\Gamma(B^-\to   F_{-1}^-  \overline \Xi_{\bar{c}}^0)
=\Gamma(\overline B^0\to   F_1^+  \bar \Xi_{\bar{c}}^- ),
\Gamma(B^-\to   F_{-3/2}^-  \overline \Xi_{\bar{c}}^0)
=\Gamma(\overline B^0_s\to   F_{-1/2}^+  \bar \Lambda_{\bar{c}}^-),\\
&\Gamma(B^-\to   F_{-3/2}^-  \overline \Sigma_{\bar{c}}^{0})
=2\Gamma(B^-\to   F_{-1}^-  \overline \Xi_{\bar{c}}^{\prime0})
=\Gamma(\overline B^0\to   F_{3/2}^{++}  \bar \Sigma_{\bar{c}}^{--})
=2\Gamma(\overline B^0\to   F_1^+  \bar \Xi_{\bar{c}}^{\prime-}),\\
&\Gamma(B^-\to   F_{-1}^-  \overline \Omega_{\bar{c}}^{0})
=2\Gamma(B^-\to   F_{-3/2}^-  \overline \Xi_{\bar{c}}^{\prime0})
=\Gamma(\overline B^0_s\to   F_{1/2}^{++}  \bar \Sigma_{\bar{c}}^{--})
=2\Gamma(\overline B^0_s\to   F_{-1/2}^+  \bar \Sigma_{\bar{c}}^{-}),\\
&\Gamma(\overline B^0_s\to   F_{-3/2}^-  \overline \Xi^{\prime+})
=\Gamma(\overline B^0\to   F_{-1}^-  \overline \Xi^{\prime+})
=\Gamma(\overline B^0\to   F_{-3/2}^-  \overline \Sigma^{\prime+})
=\frac{1}{3}\Gamma(\overline B^0_s\to   F_{-1}^-  \overline \Omega^+),\\
&\Gamma(\overline B^0_s\to   F_{-1}^-  \overline \Xi^{\prime+})
=\Gamma(\overline B^0\to   F_{-1}^-  \overline \Sigma^{\prime+})
=\frac{1}{3}\Gamma(\overline B^0\to   F_{-3/2}^-  \overline \Delta^{+})
=\Gamma(\overline B^0_s\to   F_{-3/2}^-  \overline \Sigma^{\prime+}),\\
&\Gamma(B^-\to   F_{-3/2}^-  \overline \Sigma^{\prime0})
=\Gamma(B^-\to   F_{-1}^-  \overline \Xi^{\prime0}),
\Gamma(\overline B^0_s\to   F_{1/2}^{++}  \overline \Delta^{--})
=\frac{1}{3}\Gamma(\overline B^0_s\to   F_{-1/2}^+  \overline \Delta^{-}),\\
&\Gamma(\overline B^0\to   F_1^+  \overline \Sigma^{\prime-})
=\Gamma(\overline B^0\to   F_{3/2}^{++}  \overline \Delta^{--}),
\Gamma(B^-\to   F_{-1}^-  \overline \Sigma^{\prime0})
=\frac{1}{2}\Gamma(B^-\to   F_{-3/2}^-  \overline \Delta^{0}).
\end{align*}

It can be found that for the concerned pentaquark $T_{15}$ states, the difference between the decay widths of different production processes from $B$ mesons is relatively large, but they are also interrelated each other. Once anyone decay channel will be detected in the future, we can give other decay widths. And also if two decay widths can be measured, our predictions can also be tested.
Considering the CKM and detection efficiency, one may exclude some channels with less important contribution. We remove all channels with the hadrons $\pi^0, n, \Sigma^+(\to p\pi^0), \Sigma^-(\to n\pi^-)$, $\Xi^0(\to \Lambda \pi^0)$ in the final states, but keep the processes with $\pi^{\pm}, \Sigma^0(\to N\pi\gamma)$, $\Xi^-(\to \Lambda \pi^-)$ and $\Lambda^0(\to p\pi^-)$. Therefor, the golden channels producing the singly charm pentaquark ground states are selected in Tab.~\ref{tab:goldenc}, from which, we screen out several finest processes for the concerned pentaquark $F_{15}$,
\begin{eqnarray}
\overline B^0_s\to   F_{-1/2}^+  \overline p ,
~\overline B^0\to   F_{3/2}^{++} \overline \Sigma_{\bar{c}}^{--},~
\overline B^0\to   F_{-1}^-  \overline \Xi^+  ,~
B^-\to   F_{-3/2}^-  \overline \Lambda^0,~\overline B^0\to   F_1^+ \overline \Xi_{\bar{c}}^{\prime-} ,~\overline B^0_s\to   F_1^{++}  \overline \Delta^{--} .\nonumber
\end{eqnarray}
One can make a rough estimate about the branching ratios of the processes, by assuming the weak coupling constants for the B coupling with final pentaquark and baryon\cite{Hu:2022qlr}. Immediately, we find that the branching ratios can reach to the order of $10^{-7}$.
\begin{eqnarray*}
Br(\overline B^0_s\!\!\to   F_{\frac{-1}{2}}^+  \overline p)=4.8\times10^{-9},~ Br( \overline B^0\!\!\to   F_{3/2}^{++} \overline \Sigma_{\bar{c}}^{--})=2.4\times10^{-8},
 Br(\overline B^0\!\!\to   F_{-1}^-  \overline \Xi^+)=8.2\times10^{-9},\\ Br(B^-\!\!\to   F_{\frac{-3}{2}}^-  \overline \Lambda^0)=5.6\times10^{-9},~ Br(\overline B^0\!\!\to   F_1^+ \overline \Xi_{\bar{c}}^{\prime-})=1.1\times10^{-7},~ Br(\overline B^0_s\!\!\to   F_1^{++}  \overline \Delta^{--})=5.6\times10^{-9}.
\end{eqnarray*}
Here, the weak couplings constants are near $1.0$ KeV order, and one monopole function is introduced to describe the inner structure effect of the interaction vertices, $
F({\bf q}^2)= { \Lambda^2 \over \Lambda^2 + {\bf q}^2}$. The parameter $\Lambda=300 $ MeV, and $\bf{q}^2$ is the anti-baryon three-momentum in the rest frame of the $B$ meson.
\begin{table}
\caption{The golden channels for the production of concerned ground pentaquark states from B mesons.}\label{tab:goldenc}
\begin{tabular}{cccccccc}\hline\hline
$B^-\to   F_{-1}^-  \overline \Xi_{\bar{c}}^0 $ &\quad $B^-\to   F_{-3/2}^-  \overline \Lambda^0 $&\quad $B^-\to   F_{-3/2}^-  \overline \Sigma^0 $ &\quad
$B^-\to   F_{-3/2}^-  \overline \Sigma_{\bar{c}}^{0} $\\
$B^-\to   F_{-1}^-  \overline \Xi_{\bar{c}}^{\prime0} $& $B^-\to   F_{-3/2}^-  \overline \Sigma^{\prime0}$ &$B^-\to   F_{-1}^-  \overline \Xi^{\prime0}$ \\\hline
$\overline B^0\to   F_{-1}^-  \overline \Xi^+ $&$\overline B^0\to   F_{3/2}^{++} \overline \Sigma_{\bar{c}}^{--} $&\quad$\overline B^0\to   F_1^+ \overline \Xi_{\bar{c}}^{\prime-} $&$\overline B^0\to   F_{-1}^-  \overline \Xi^{\prime+} $\\
$\overline B^0\to   F_{-3/2}^-  \overline \Sigma^{\prime+} $\\\hline
$\overline B^0_s\to   F_{-1/2}^+  \overline p $&\quad$\overline B^0_s\to   F_{-3/2}^-  \overline \Xi^+ $&$\overline B^0_s\to   F_{-3/2}^-  \overline \Xi^{\prime+} $&$\overline B^0_s\to   F_{-1}^-  \overline \Omega^+ $\\
$\overline B^0_s\to   F_{1/2}^{++}  \overline \Delta^{--} $&$\overline B^0_s\to   F_{-1/2}^+  \overline \Delta^{-} $
\\\hline \hline
\end{tabular}
\end{table}
\begin{table}
\caption{The productions of concerned pentaquark 15 state and light anti-baryons $({qqq})_{8/10}$ or anti-charm anti-baryon $(\bar{c}qq)_{3/\bar 6}$ from B mesons.}\label{tab:F15concern}\begin{tabular}{|cc|cc|cc|c|c|c|c|c}\hline\hline
channel & amplitude& channel & amplitude&channel & amplitude\\\hline
$B^-\to   F_{-3/2}^-  \overline \Xi_{\bar{c}}^0 $ & $ a_1 V_{cd}^*$&$B^-\to   F_{-1}^-  \overline \Xi_{\bar{c}}^0 $ & $ a_1 V_{cs}^*$&
$\overline B^0\to   F_1^+  \bar \Xi_{\bar{c}}^- $ & $ -a_1 V_{cs}^*$\\
$\overline B^0_s\to   F_{-1/2}^+  \bar \Lambda_{\bar{c}}^- $ & $ a_1 V_{cd}^*$&&&&\\\hline\hline
$B^-\to   F_{-3/2}^-  \overline \Sigma_{\bar{c}}^{0} $ & $ -b_1 V_{cs}^*$&
$B^-\to   F_{-3/2}^-  \overline \Xi_{\bar{c}}^{\prime0} $ & $ \frac{b_1 V_{cd}^*}{\sqrt{2}}$&
$B^-\to   F_{-1}^-  \overline \Xi_{\bar{c}}^{\prime0} $ & $ \frac{b_1 V_{cs}^*}{-\sqrt{2}}$\\
$B^-\to   F_{-1}^-  \overline \Omega_{\bar{c}}^{0} $ & $ b_1 V_{cd}^*$&
$\overline B^0\to   F_{3/2}^{++} \overline \Sigma_{\bar{c}}^{--} $ & $ b_1 V_{cs}^*$&
$\overline B^0\to   F_1^+ \overline \Xi_{\bar{c}}^{\prime-} $ & $ \frac{b_1 V_{cs}^*}{\sqrt{2}}$\\
$\overline B^0_s\to   F_{1/2}^{++}  \overline \Sigma_{\bar{c}}^{--} $ & $ -b_1 V_{cd}^*$&
$\overline B^0_s\to   F_{-1/2}^+  \overline \Sigma_{\bar{c}}^{-} $ & $ \frac{b_1 V_{cd}^*}{-\sqrt{2}}$&&\\\hline\hline
$B^-\to   F_{-3/2}^-  \overline \Lambda^0 $ & $ \frac{\left(c_2+c_3+c_4\right) V_{ud}^*}{\sqrt{6}}$&
$B^-\to   F_{-3/2}^-  \overline \Sigma^0 $ & $ \frac{\left(c_2+c_3-c_4\right) V_{ud}^*}{\sqrt{2}}$&
$B^-\to   F_{-3/2}^-  \overline n $ & $ c_4 V_{us}^*$\\
$B^-\to   F_{-1}^-  \overline \Lambda^0 $ & $ \frac{\left(c_2+c_3-2 c_4\right) V_{us}^*}{\sqrt{6}}$&
$B^-\to   F_{-1}^-  \overline \Sigma^0 $ & $ \frac{\left(c_2+c_3\right) V_{us}^*}{\sqrt{2}}$&
$B^-\to   F_{-1}^-  \overline \Xi^0 $ & $ c_4 V_{ud}^*$\\
$\overline B^0\to   F_{-3/2}^-  \overline \Sigma^+ $ & $ \left(c_1+c_3\right) V_{ud}^*$&
$\overline B^0\to   F_1^+  \overline \Sigma^- $ & $ c_2 V_{us}^*$&
$\overline B^0\to   F_{-1}^-  \overline \Sigma^+ $ & $ c_3 V_{us}^*$\\
$\overline B^0\to   F_{-1}^-  \overline \Xi^+ $ & $ c_1 V_{ud}^*$&
$\overline B^0_s\to   F_{-1/2}^+  \overline p $ & $ c_2 V_{ud}^*$&
$\overline B^0_s\to   F_{-3/2}^-  \overline \Sigma^+ $ & $ c_1V_{us}^*$\\
$\overline B^0_s\to   F_{-3/2}^-  \overline \Xi^+ $ & $ c_3 V_{ud}^*$&
$\overline B^0_s\to   F_{-1}^-  \overline \Xi^+ $ & $ \left(c_1+c_3\right) V_{us}^*$&&\\\hline
$\overline B^0\to   F_1^+  \overline \Sigma^{\prime-} $ & $ \frac{d_1 V_{us}^*}{\sqrt{3}}$
&$\overline B^0\to   F_{-1}^-  \overline \Sigma^{\prime+} $ & $ \frac{d_2 V_{us}^*}{-\sqrt{3}}$&
$\overline B^0_s\to   F_{-1/2}^+  \overline \Delta^{-} $ & $ \frac{d_1 V_{ud}^*}{-\sqrt{3}}$\\
$\overline B^0_s\to   F_{-3/2}^-  \overline \Xi^{\prime+} $ & $ \frac{d_2 V_{ud}^*}{\sqrt{3}}$&
$\overline B^0_s\to   F_{-1}^-  \overline \Omega^+ $ & $ d_2 V_{ud}^*$&
$B^-\to   F_{-3/2}^-  \overline \Delta^{0} $ & $ \frac{\left(d_1+d_2\right) V_{us}^*}{-\sqrt{3}}$\\
$B^-\to   F_{-1}^-  \overline \Sigma^{\prime0} $ & $ \frac{\left(d_1+d_2\right) V_{us}^*}{-\sqrt{6}}$&
$\overline B^0\to   F_{3/2}^{++}  \overline \Delta^{--} $ & $ d_1 V_{us}^*$&
$\overline B^0\to   F_{-3/2}^-  \overline \Delta^{+} $ & $ -d_2 V_{us}^*$\\
$\overline B^0\to   F_{-1}^-  \overline \Xi^{\prime+} $ & $ \frac{d_2 V_{ud}^*}{\sqrt{3}}$&
$B^-\to   F_{-3/2}^-  \overline \Sigma^{\prime0} $ & $ \frac{\left(d_1+d_2\right) V_{ud}^*}{\sqrt{6}}$&
$B^-\to   F_{-1}^-  \overline \Xi^{\prime0} $ & $ \frac{\left(d_1+d_2\right) V_{ud}^*}{\sqrt{3}}$\\
$\overline B^0\to   F_{-3/2}^-  \overline \Sigma^{\prime+} $ & $ \frac{d_2 V_{ud}^*}{\sqrt{3}}$&
$\overline B^0_s\to   F_{1/2}^{++}  \overline \Delta^{--} $ & $ -d_1 V_{ud}^*$&
$\overline B^0_s\to   F_{-3/2}^-  \overline \Sigma^{\prime+} $ & $ \frac{d_2 V_{us}^*}{-\sqrt{3}}$\\
$\overline B^0_s\to   F_{-1}^-  \overline \Xi^{\prime+} $ & $ \frac{d_2 V_{us}^*}{-\sqrt{3}}$&&&&\\\hline

\end{tabular}
\end{table}
\section{Conclusions}
In this work, we study the mass splitting of the S-wave singly charm pentaquark state ${c\bar qqqq}(q=u,d,s)$ in the framework of non-relativity doubly heavy triquark-diquark model, in which the hyperfine structure from spin-spin and spin-orbit interaction. We find that  $c\bar u sdd, \ c\bar d suu,\ c\bar d uud$ and $c \bar s udd$ with parity $\frac{1}{2}^-$ are below their strong decay thresholds, which imply that they are the stable states. It should be checked in future experiments.

Within the SU(3) flavor symmetry, we discuss the production of the ground pentaquark state from B meson, several golden channels are selected, in particular, the estimation of branching ratios can reach to a sizeable order of $10^{-7}$. We further intend to consider the processes with more effective means, such as the effective Lagrangian method, expecting to acquire more convincing results for the experimental detection.

\appendix
\section{tensor decompositon}\label{sec:tensorform}
We list the tensor representations of pentaquark $c \bar q qqq$ with $6$, $15$ and $24$ respectively. The 6 states $T_6$ can be obtained from the decomposition of $1\otimes 6$, $8\otimes 3$ or $8\otimes 6$.
\begin{small}
\begin{align*}
&(T_6)_{\{11\}} = -c\bar{d}\text{duu}+c\bar{d}\text{udu}-c\bar{s}\text{suu}+c\bar{s}\text{usu},\\&
(T_6)_{\{12\}} = \frac{1}{2} (-c\bar{d}\text{dud}+c\bar{d}\text{udd}+c\bar{s}\text{dsu}-c\bar{s}\text{sdu}-c\bar{s}\text{sud}+c\bar{s}\text{usd}+c\bar{u}\text{duu}-c\bar{u}\text{udu}),\\&
(T_6)_{\{13\}} = \frac{1}{2} (-c\bar{d}\text{dsu}-c\bar{d}\text{dus}+c\bar{d}\text{sdu}+c\bar{d}\text{uds}-c\bar{s}\text{sus}+c\bar{s}\text{uss}+c\bar{u}\text{suu}-c\bar{u}\text{usu}),\\&
(T_6)_{\{22\}} = c\bar{s}\text{dsd}-c\bar{s}\text{sdd}+c\bar{u}\text{dud}-c\bar{u}\text{udd},\\&
(T_6)_{\{23\}} = \frac{1}{2} (-c\bar{d}\text{dsd}+c\bar{d}\text{sdd}+c\bar{s}\text{dss}-c\bar{s}\text{sds}+c\bar{u}\text{dus}+c\bar{u}\text{sud}-c\bar{u}\text{uds}-c\bar{u}\text{usd}),\\&
(T_6)_{\{33\}} = -c\bar{d}\text{dss}+c\bar{d}\text{sds}+c\bar{u}\text{sus}-c\bar{u}\text{uss}.
\end{align*}
\end{small}

And the $15$ states $T_{15}$ are decomposed from $8\otimes 3$ or $8\otimes 6$,
\begin{small}
\begin{align*}
&(T_{15})^{\{11\}}_1 = \frac{1}{2} (c\bar{d}\text{dsd}-c\bar{d}\text{sdd}+c\bar{s}\text{dss}-c\bar{s}\text{sds}+2 c\bar{u}\text{dsu}+c\bar{u}\text{dus}-2 c\bar{u}\text{sdu}-c\bar{u}\text{sud}-c\bar{u}\text{uds}+c\bar{u}\text{usd}) ,\\&
(T_{15})^{\{11\}}_2 = c\bar{u}\text{dsd}-c\bar{u}\text{sdd},
 (T_{15})^{\{11\}}_3 = c\bar{u}\text{dss}-c\bar{u}\text{sds} ,\\&
(T_{15})^{\{12\}}_1 = \frac{1}{4} (c\bar{d}\text{dsu}+c\bar{d}\text{dus}-c\bar{d}\text{sdu}-c\bar{d}\text{uds}+c\bar{s}\text{sus}-c\bar{s}\text{uss}+3 c\bar{u}\text{suu}-3 c\bar{u}\text{usu}) ,\\&
(T_{15})^{\{12\}}_2 = \frac{1}{4} (3 c\bar{d}\text{dsd}-3 c\bar{d}\text{sdd}+c\bar{s}\text{dss}-c\bar{s}\text{sds}+c\bar{u}\text{dus}+c\bar{u}\text{sud}-c\bar{u}\text{uds}-c\bar{u}\text{usd}) ,\\&
(T_{15})^{\{12\}}_3 = \frac{1}{2} (c\bar{d}\text{dss}-c\bar{d}\text{sds}+c\bar{u}\text{sus}-c\bar{u}\text{uss}) ,\\&
(T_{15})^{\{13\}}_1 = \frac{1}{4} (-c\bar{d}\text{dud}+c\bar{d}\text{udd}+c\bar{s}\text{dsu}-c\bar{s}\text{sdu}-c\bar{s}\text{sud}+c\bar{s}\text{usd}-3 c\bar{u}\text{duu}+3 c\bar{u}\text{udu}) ,\\&
(T_{15})^{\{13\}}_2 = \frac{1}{2} (c\bar{s}\text{dsd}-c\bar{s}\text{sdd}-c\bar{u}\text{dud}+c\bar{u}\text{udd}) ,\\&
(T_{15})^{\{13\}}_3 = \frac{1}{4} (c\bar{d}\text{dsd}-c\bar{d}\text{sdd}+3 c\bar{s}\text{dss}-3 c\bar{s}\text{sds}-c\bar{u}\text{dus}-c\bar{u}\text{sud}+c\bar{u}\text{uds}+c\bar{u}\text{usd}) ,\\&
(T_{15})^{\{22\}}_1 = c\bar{d}\text{suu}-c\bar{d}\text{usu} ,\\&
(T_{15})^{\{22\}}_2 = \frac{1}{2} (-c\bar{d}\text{dsu}+c\bar{d}\text{dus}+c\bar{d}\text{sdu}+2 c\bar{d}\text{sud}-c\bar{d}\text{uds}-2 c\bar{d}\text{usd}+c\bar{s}\text{sus}-c\bar{s}\text{uss}+c\bar{u}\text{suu}-c\bar{u}\text{usu}) ,\\&
(T_{15})^{\{22\}}_3 = c\bar{d}\text{sus}-c\bar{d}\text{uss} ,
(T_{15})^{\{23\}}_1 = \frac{1}{2} (-c\bar{d}\text{duu}+c\bar{d}\text{udu}+c\bar{s}\text{suu}-c\bar{s}\text{usu}) ,\\&
(T_{15})^{\{23\}}_2 = \frac{1}{4} (-3 c\bar{d}\text{dud}+3 c\bar{d}\text{udd}-c\bar{s}\text{dsu}+c\bar{s}\text{sdu}+c\bar{s}\text{sud}-c\bar{s}\text{usd}-c\bar{u}\text{duu}+c\bar{u}\text{udu}) ,\\&
(T_{15})^{\{23\}}_3 = \frac{1}{4} (-c\bar{d}\text{dsu}-c\bar{d}\text{dus}+c\bar{d}\text{sdu}+c\bar{d}\text{uds}+3 c\bar{s}\text{sus}-3 c\bar{s}\text{uss}+c\bar{u}\text{suu}-c\bar{u}\text{usu}) ,\\&
(T_{15})^{\{33\}}_1 = c\bar{s}\text{udu}-c\bar{s}\text{duu} ,
(T_{15})^{\{33\}}_2 = c\bar{s}\text{udd}-c\bar{s}\text{dud} ,\\&
(T_{15})^{\{33\}}_3 = \frac{1}{2} (-c\bar{d}\text{dud}+c\bar{d}\text{udd}-c\bar{s}\text{dsu}-2 c\bar{s}\text{dus}+c\bar{s}\text{sdu}-c\bar{s}\text{sud}+2 c\bar{s}\text{uds}+c\bar{s}\text{usd}-c\bar{u}\text{duu}+c\bar{u}\text{udu}).
\end{align*}
\end{small}
In addition, the tensor $T_{24}$ is derived from $8\otimes 6$.
\begin{small}
\begin{align*}
&(T_{24})_{1}^{\{111\}} = \frac{1}{5} (-c\bar{d}\text{duu}-c\bar{d}\text{udu}-c\bar{d}\text{uud}-c\bar{s}\text{suu}-c\bar{s}\text{usu}-c\bar{s}\text{uus}+2 c\bar{u}\text{uuu}) ,\\&
(T_{24})_{1}^{\{112\}} = \frac{1}{15} (-2 c\bar{d}\text{ddu}-2 c\bar{d}\text{dud}-2 c\bar{d}\text{udd}-c\bar{s}\text{dsu}-c\bar{s}\text{dus}-c\bar{s}\text{sdu}-c\bar{s}\text{sud}-c\bar{s}\text{uds}-c\bar{s}\text{usd}+3 c\bar{u}\text{duu}\\&+3 c\bar{u}\text{udu}+3 c\bar{u}\text{uud}) ,\\&
(T_{24})_{1}^{\{113\}} = \frac{1}{15} (-c\bar{d}\text{dsu}-c\bar{d}\text{dus}-c\bar{d}\text{sdu}-c\bar{d}\text{sud}-c\bar{d}\text{uds}-c\bar{d}\text{usd}-2 c\bar{s}\text{ssu}-2 c\bar{s}\text{sus}-2 c\bar{s}\text{uss}+3 c\bar{u}\text{suu}\\&+3 c\bar{u}\text{usu} +3 c\bar{u}\text{uus}) ,\\&
(T_{24})_{1}^{\{122\}}= \frac{1}{15} (-3 c\bar{d}\text{ddd}-c\bar{s}\text{dds}-c\bar{s}\text{dsd}-c\bar{s}\text{sdd}+4 c\bar{u}\text{ddu}+4 c\bar{u}\text{dud}+4 c\bar{u}\text{udd}) ,\\&
(T_{24})_{1}^{\{123\}} = \frac{1}{15} (-c\bar{d}\text{dds}-c\bar{d}\text{dsd}-c\bar{d}\text{sdd}-c\bar{s}\text{dss}-c\bar{s}\text{sds}-c\bar{s}\text{ssd}+2 c\bar{u}\text{dsu}+2 c\bar{u}\text{dus}+2 c\bar{u}\text{sdu}+2 c\bar{u}\text{sud}\\&+2 c\bar{u}\text{uds}+2 c\bar{u}\text{usd}) ,\\&
(T_{24})_{1}^{\{133\}} = \frac{1}{15} (-c\bar{d}\text{dss}-c\bar{d}\text{sds}-c\bar{d}\text{ssd}-3 c\bar{s}\text{sss}+4 c\bar{u}\text{ssu}+4 c\bar{u}\text{sus}+4 c\bar{u}\text{uss}) ,\\&
(T_{24})_{1}^{\{222\}} = c\bar{u}\text{ddd} ,
(T_{24})_{1}^{\{223\}} = \frac{1}{3} (c\bar{u}\text{dds}+c\bar{u}\text{dsd}+c\bar{u}\text{sdd}) ,
(T_{24})_{1}^{\{233\}} = \frac{1}{3} (c\bar{u}\text{dss}+c\bar{u}\text{sds}+c\bar{u}\text{ssd}) ,\\&
(T_{24})_{1}^{\{332\}} = \frac{1}{3} (c\bar{u}\text{dss}+c\bar{u}\text{sds}+c\bar{u}\text{ssd}) ,
(T_{24})_{1}^{\{333\}} = c\bar{u}\text{sss} ,
(T_{24})_{2}^{\{111\}} = c\bar{d}\text{uuu} ,\\&
(T_{24})_{2}^{\{112\}} = \frac{1}{15} (4 c\bar{d}\text{duu}+4 c\bar{d}\text{udu}+4 c\bar{d}\text{uud}-c\bar{s}\text{suu}-c\bar{s}\text{usu}-c\bar{s}\text{uus}-3 c\bar{u}\text{uuu}) ,\\&
(T_{24})_{2}^{\{113\}} = \frac{1}{3} (c\bar{d}\text{suu}+c\bar{d}\text{usu}+c\bar{d}\text{uus}) ,\\&
(T_{24})_{2}^{\{122\}} = \frac{1}{15} (3 c\bar{d}\text{ddu}+3 c\bar{d}\text{dud}+3 c\bar{d}\text{udd}-c\bar{s}\text{dsu}-c\bar{s}\text{dus}-c\bar{s}\text{sdu}-c\bar{s}\text{sud}-c\bar{s}\text{uds}-c\bar{s}\text{usd}-2 c\bar{u}\text{duu}\\&-2 c\bar{u}\text{udu}-2 c\bar{u}\text{uud}) ,\\&
(T_{24})_{2}^{\{123\}} = \frac{1}{15} (2 c\bar{d}\text{dsu}+2 c\bar{d}\text{dus}+2 c\bar{d}\text{sdu}+2 c\bar{d}\text{sud}+2 c\bar{d}\text{uds}+2 c\bar{d}\text{usd}-c\bar{s}\text{ssu}-c\bar{s}\text{sus}-c\bar{s}\text{uss}-c\bar{u}\text{suu}\\&-c\bar{u}\text{usu}-c\bar{u}\text{uus}) ,\\&
(T_{24})_{2}^{\{133\}} = \frac{1}{3} (c\bar{d}\text{ssu}+c\bar{d}\text{sus}+c\bar{d}\text{uss}) ,\\&
(T_{24})_{2}^{\{222\}} = \frac{1}{5} (2 c\bar{d}\text{ddd}-c\bar{s}\text{dds}-c\bar{s}\text{dsd}-c\bar{s}\text{sdd}-c\bar{u}\text{ddu}-c\bar{u}\text{dud}-c\bar{u}\text{udd}) ,\\&
(T_{24})_{2}^{\{223\}} = \frac{1}{15} (3 c\bar{d}\text{dds}+3 c\bar{d}\text{dsd}+3 c\bar{d}\text{sdd}-2 c\bar{s}\text{dss}-2 c\bar{s}\text{sds}-2 c\bar{s}\text{ssd}-c\bar{u}\text{dsu}-c\bar{u}\text{dus}-c\bar{u}\text{sdu}-c\bar{u}\text{sud}\\&-c\bar{u}\text{uds}-c\bar{u}\text{usd}) ,\\&
(T_{24})_{2}^{\{233\}} = \frac{1}{15} (4 c\bar{d}\text{dss}+4 c\bar{d}\text{sds}+4 c\bar{d}\text{ssd}-3 c\bar{s}\text{sss}-c\bar{u}\text{ssu}-c\bar{u}\text{sus}-c\bar{u}\text{uss}) ,\\&
(T_{24})_{2}^{\{333\}} = c\bar{d}\text{sss} ,
(T_{24})_{3}^{\{111\}} = c\bar{s}\text{uuu} ,
(T_{24})_{3}^{\{112\}} = \frac{1}{3} (c\bar{s}\text{duu}+c\bar{s}\text{udu}+c\bar{s}\text{uud}) ,\\&
(T_{24})_{3}^{\{113\}} = \frac{1}{15} (-c\bar{d}\text{duu}-c\bar{d}\text{udu}-c\bar{d}\text{uud}+4 c\bar{s}\text{suu}+4 c\bar{s}\text{usu}+4 c\bar{s}\text{uus}-3 c\bar{u}\text{uuu}) ,\\&
(T_{24})_{3}^{\{122\}} = \frac{1}{3} (c\bar{s}\text{ddu}+c\bar{s}\text{dud}+c\bar{s}\text{udd}) ,\\&
(T_{24})_{3}^{\{123\}} = \frac{1}{15} (-c\bar{d}\text{ddu}-c\bar{d}\text{dud}-c\bar{d}\text{udd}+2 c\bar{s}\text{dsu}+2 c\bar{s}\text{dus}+2 c\bar{s}\text{sdu}+2 c\bar{s}\text{sud}+2 c\bar{s}\text{uds}+2 c\bar{s}\text{usd}-c\bar{u}\text{duu}\\&-c\bar{u}\text{udu}-c\bar{u}\text{uud}) ,\\&
(T_{24})_{3}^{\{133\}} = \frac{1}{15} (-c\bar{d}\text{dsu}-c\bar{d}\text{dus}-c\bar{d}\text{sdu}-c\bar{d}\text{sud}-c\bar{d}\text{uds}-c\bar{d}\text{usd}+3 c\bar{s}\text{ssu}+3 c\bar{s}\text{sus}+3 c\bar{s}\text{uss}-2 c\bar{u}\text{suu}\\&-2 c\bar{u}\text{usu}-2 c\bar{u}\text{uus}) ,\\&
(T_{24})_{3}^{\{222\}} = c\bar{s}\text{ddd} ,
(T_{24})_{3}^{\{223\}} = \frac{1}{15} (-3 c\bar{d}\text{ddd}+4 c\bar{s}\text{dds}+4 c\bar{s}\text{dsd}+4 c\bar{s}\text{sdd}-c\bar{u}\text{ddu}-c\bar{u}\text{dud}-c\bar{u}\text{udd}) ,\\&
(T_{24})_{3}^{\{233\}} = \frac{1}{15} (-2 c\bar{d}\text{dds}-2 c\bar{d}\text{dsd}-2 c\bar{d}\text{sdd}+3 c\bar{s}\text{dss}+3 c\bar{s}\text{sds}+3 c\bar{s}\text{ssd}-c\bar{u}\text{dsu}-c\bar{u}\text{dus}-c\bar{u}\text{sdu}-c\bar{u}\text{sud}\\&-c\bar{u}\text{uds}-c\bar{u}\text{usd}) ,\\&
(T_{24})_{3}^{\{333\}} = \frac{1}{5} (-c\bar{d}\text{dss}-c\bar{d}\text{sds}-c\bar{d}\text{ssd}+2 c\bar{s}\text{sss}-c\bar{u}\text{ssu}-c\bar{u}\text{sus}-c\bar{u}\text{uss}).
\end{align*}
\end{small}
\section{states $6$ and $15$}\label{sec:tables}
\begin{table}
\caption{The productions of pentaquark $F_{15}$ and light anti-baryons $({qqq})_{8/10}$ from B mesons.}
\label{tab:P15full1}\begin{tabular}{|cc|cc|cc|c|c}\hline\hline
channel & amplitude & channel & amplitude &channel & amplitude\\\hline
$B^-\to   F_{-1}^0  \bar \Lambda_{\bar{c}}^- $ & $ a_1 V_{cd}^*$&
$B^-\to   F_{-1/2}^0  \bar \Xi_{\bar{c}}^- $ & $ -a_1 V_{cd}^*$&
$B^-\to   {F'}_{-1/2}^0  \bar \Lambda_{\bar{c}}^- $ & $ a_1 V_{cs}^*$\\
$B^-\to   F_{-3/2}^-  \overline \Xi_{\bar{c}}^0 $ & $ a_1 V_{cd}^*$&
$B^-\to   F_0^0  \bar \Xi_{\bar{c}}^- $ & $ -a_1 V_{cs}^*$&
$B^-\to   F_{-1}^-  \overline \Xi_{\bar{c}}^0 $ & $ a_1 V_{cs}^*$\\
$\overline B^0\to   {F'}_0^+  \bar \Lambda_{\bar{c}}^- $ & $ a_1 V_{cd}^*$&
$\overline B^0\to   F_{1/2}^+  \bar \Xi_{\bar{c}}^- $ & $ a_1 V_{cd}^*$&
$\overline B^0\to   {F'}_{1/2}^+  \bar \Lambda_{\bar{c}}^- $ & $ a_1 V_{cs}^*$\\
$\overline B^0\to   {F'}_{1/2}^+  \bar \Xi_{\bar{c}}^- $ & $ a_1 V_{cd}^*$&
$\overline B^0\to   F_{-1/2}^0  \overline \Xi_{\bar{c}}^0 $ & $ a_1 V_{cd}^*$&
$\overline B^0\to   F_1^+  \bar \Xi_{\bar{c}}^- $ & $ -a_1 V_{cs}^*$\\
$\overline B^0\to   F_0^0  \overline \Xi_{\bar{c}}^0 $ & $ a_1 V_{cs}^*$&
$\overline B^0_s\to   F_{-1/2}^+  \bar \Lambda_{\bar{c}}^- $ & $ a_1 V_{cd}^*$&
$\overline B^0_s\to   {F'}_0^+  \bar \Lambda_{\bar{c}}^- $ & $ -a_1 V_{cs}^*$\\
$\overline B^0_s\to   {F'}_0^+  \bar \Xi_{\bar{c}}^- $ & $ -a_1 V_{cd}^*$&
$\overline B^0_s\to   F_0^+  \bar \Lambda_{\bar{c}}^- $ & $ -a_1 V_{cs}^*$&
$\overline B^0_s\to   F_{-1}^0  \overline \Xi_{\bar{c}}^0 $ & $ a_1 V_{cd}^*$\\
$\overline B^0_s\to   {F'}_{1/2}^+  \bar \Xi_{\bar{c}}^- $ & $ -a_1 V_{cs}^*$&
$\overline B^0_s\to   {F'}_{-1/2}^0  \overline \Xi_{\bar{c}}^0 $ & $ a_1 V_{cs}^*$&&\\\hline
$B^-\to   F_0^+  \bar \Sigma_{\bar{c}}^{--} $ & $ -b_1 V_{cd}^*$&
$B^-\to   F_{-1}^0  \bar \Sigma_{\bar{c}}^{-} $ & $ \frac{b_1 V_{cd}^*}{-\sqrt{2}}$&
$B^-\to   F_{1/2}^+  \bar \Sigma_{\bar{c}}^{--} $ & $ b_1 V_{cs}^*$\\
$B^-\to   F_{-1/2}^0  \bar \Sigma_{\bar{c}}^{-} $ & $ \sqrt{2} b_1 V_{cs}^*$&
$B^-\to   F_{-1/2}^0  \bar \Xi_{\bar{c}}^{\prime-} $ & $ \frac{b_1 V_{cd}^*}{-\sqrt{2}}$&
$B^-\to   {F'}_{-1/2}^0  \bar \Sigma_{\bar{c}}^{-} $ & $ \frac{b_1 V_{cs}^*}{\sqrt{2}}$\\
$B^-\to   {F'}_{-1/2}^0  \bar \Xi_{\bar{c}}^{\prime-} $ & $ -\sqrt{2} b_1 V_{cd}^*$&
$B^-\to   F_{-3/2}^-  \overline \Sigma_{\bar{c}}^{0} $ & $ -b_1 V_{cs}^*$&
$B^-\to   F_{-3/2}^-  \overline \Xi_{\bar{c}}^{\prime0} $ & $ \frac{b_1 V_{cd}^*}{\sqrt{2}}$\\
$B^-\to   F_0^0  \bar \Xi_{\bar{c}}^{\prime-} $ & $ \frac{b_1 V_{cs}^*}{\sqrt{2}}$&
$B^-\to   F_{-1}^-  \overline \Xi_{\bar{c}}^{\prime0} $ & $ \frac{b_1 V_{cs}^*}{-\sqrt{2}}$&
$B^-\to   F_{-1}^-  \overline \Omega_{\bar{c}}^{0} $ & $ b_1 V_{cd}^*$\\
$\overline B^0\to   F_1^{++}  \bar \Sigma_{\bar{c}}^{--} $ & $ -b_1 V_{cd}^*$&
$\overline B^0\to   {F'}_0^+  \bar \Sigma_{\bar{c}}^{-} $ & $ \frac{b_1 V_{cd}^*}{-\sqrt{2}}$&
$\overline B^0\to   F_{3/2}^{++}  \bar \Sigma_{\bar{c}}^{--} $ & $ b_1 V_{cs}^*$\\
$\overline B^0\to   F_{1/2}^+  \bar \Sigma_{\bar{c}}^{-} $ & $ -\sqrt{2} b_1 V_{cs}^*$&
$\overline B^0\to   F_{1/2}^+  \bar \Xi_{\bar{c}}^{\prime-} $ & $ \frac{b_1 V_{cd}^*}{\sqrt{2}}$&
$\overline B^0\to   {F'}_{1/2}^+  \bar \Sigma_{\bar{c}}^{-} $ & $ \frac{b_1 V_{cs}^*}{-\sqrt{2}}$\\
$\overline B^0\to   {F'}_{1/2}^+  \bar \Xi_{\bar{c}}^{\prime-} $ & $ \frac{b_1 V_{cd}^*}{-\sqrt{2}}$&
$\overline B^0\to   F_{-1/2}^0  \overline \Sigma_{\bar{c}}^{0} $ & $ -b_1 V_{cs}^*$&
$\overline B^0\to   F_{-1/2}^0  \overline \Xi_{\bar{c}}^{\prime0} $ & $ \frac{b_1 V_{cd}^*}{\sqrt{2}}$\\
$\overline B^0\to   F_1^+  \bar \Xi_{\bar{c}}^{\prime-} $ & $ \frac{b_1 V_{cs}^*}{\sqrt{2}}$&
$\overline B^0\to   F_0^0  \overline \Xi_{\bar{c}}^{\prime0} $ & $ \frac{b_1 V_{cs}^*}{-\sqrt{2}}$&
$\overline B^0\to   F_0^0  \overline \Omega_{\bar{c}}^{0} $ & $ b_1 V_{cd}^*$\\
$\overline B^0_s\to   F_{1/2}^{++}  \bar \Sigma_{\bar{c}}^{--} $ & $ -b_1 V_{cd}^*$&
$\overline B^0_s\to   F_{-1/2}^+  \bar \Sigma_{\bar{c}}^{-} $ & $ \frac{b_1 V_{cd}^*}{-\sqrt{2}}$&
$\overline B^0_s\to   F_1^{++}  \bar \Sigma_{\bar{c}}^{--} $ & $ b_1 V_{cs}^*$\\
$\overline B^0_s\to   {F'}_0^+  \bar \Sigma_{\bar{c}}^{-} $ & $ \frac{b_1 V_{cs}^*}{\sqrt{2}}$&
$\overline B^0_s\to   {F'}_0^+  \bar \Xi_{\bar{c}}^{\prime-} $ & $ \frac{b_1 V_{cd}^*}{\sqrt{2}}$&
$\overline B^0_s\to   F_0^+  \bar \Sigma_{\bar{c}}^{-} $ & $ \frac{b_1 V_{cs}^*}{-\sqrt{2}}$\\
$\overline B^0_s\to   F_0^+  \bar \Xi_{\bar{c}}^{\prime-} $ & $ \sqrt{2} b_1 V_{cd}^*$&
$\overline B^0_s\to   F_{-1}^0  \overline \Sigma_{\bar{c}}^{0} $ & $ -b_1 V_{cs}^*$&
$\overline B^0_s\to   F_{-1}^0  \overline \Xi_{\bar{c}}^{\prime0} $ & $ \frac{b_1 V_{cd}^*}{\sqrt{2}}$\\
$\overline B^0_s\to   {F'}_{1/2}^+  \bar \Xi_{\bar{c}}^{\prime-} $ & $ \frac{b_1 V_{cs}^*}{\sqrt{2}}$&
$\overline B^0_s\to   {F'}_{-1/2}^0  \overline \Xi_{\bar{c}}^{\prime0} $ & $ \frac{b_1 V_{cs}^*}{-\sqrt{2}}$&
$\overline B^0_s\to   {F'}_{-1/2}^0  \overline \Omega_{\bar{c}}^{0} $ & $ b_1 V_{cd}^*$\\\hline
\end{tabular}
\end{table}
\begin{table}
\caption{The productions of pentaquark $F_{15}$ and anti-charm anti-baryons $({cqq})_{3/\bar 6}$ from B mesons.}
\label{tab:P15full2}\begin{tabular}{|cc|cc|cc|c|c}\hline\hline
channel & amplitude & channel & amplitude &channel & amplitude\\\hline
$B^-\to   F_{-1}^0  \overline p $ & $ \left(c_2+c_3\right) V_{ud}^*$&
$B^-\to   F_{-1/2}^0  \overline \Sigma^- $ & $ \left(c_2+c_3-c_4\right) V_{ud}^*$&
$B^-\to   F_{-1/2}^0  \overline p $ & $ -c_4 V_{us}^*$\\
$B^-\to   {F'}_{-1/2}^0  \overline \Sigma^- $ & $ -c_4 V_{ud}^*$&
$B^-\to   {F'}_{-1/2}^0  \overline p $ & $ \left(c_2+c_3-c_4\right) V_{us}^*$&
$B^-\to   F_{-3/2}^-  \overline \Lambda^0 $ & $ \frac{\left(c_2+c_3+c_4\right) V_{ud}^*}{\sqrt{6}}$\\
$B^-\to   F_{-3/2}^-  \overline \Sigma^0 $ & $ \frac{\left(c_2+c_3-c_4\right) V_{ud}^*}{\sqrt{2}}$&
$B^-\to   F_{-3/2}^-  \overline n $ & $ c_4 V_{us}^*$&
$B^-\to   F_0^0  \overline \Sigma^- $ & $ \left(c_2+c_3\right) V_{us}^*$\\
$B^-\to   F_{-1}^-  \overline \Lambda^0 $ & $ \frac{\left(c_2+c_3-2 c_4\right) V_{us}^*}{\sqrt{6}}$&
$B^-\to   F_{-1}^-  \overline \Sigma^0 $ & $ \frac{\left(c_2+c_3\right) V_{us}^*}{\sqrt{2}}$&
$B^-\to   F_{-1}^-  \overline \Xi^0 $ & $ c_4 V_{ud}^*$\\
$\overline B^0\to   {F'}_0^+  \overline p $ & $ c_2 V_{ud}^*$&
$\overline B^0\to   F_0^+  \overline p $ & $ c_1 V_{ud}^*$&
$\overline B^0\to   F_{-1}^0  \overline n $ & $ \left(c_1+c_3\right) V_{ud}^*$\\
$\overline B^0\to   F_{1/2}^+  \overline \Sigma^- $ & $ \left(c_1-c_2+c_4\right) V_{ud}^*$&
$\overline B^0\to   F_{1/2}^+  \overline p $ & $ c_4 V_{us}^*$&
$\overline B^0\to   {F'}_{1/2}^+  \overline \Sigma^- $ & $ -c_2 V_{ud}^*$\\
$\overline B^0\to   {F'}_{1/2}^+  \overline p $ & $ c_2 V_{us}^*$&
$\overline B^0\to   F_{-1/2}^0  \overline \Lambda^0 $ & $ \frac{\left(c_2+c_3+c_4\right)V_{ud}^*}{\sqrt{6}}$&
$\overline B^0\to   F_{-1/2}^0  \overline \Sigma^0 $ & $ \frac{\left(2 c_1-c_2+c_3+c_4\right) V_{ud}^*}{-\sqrt{2}}$\\
$\overline B^0\to   F_{-1/2}^0  \overline n $ & $ c_4 V_{us}^*$&
$\overline B^0\to   {F'}_{-1/2}^0  \overline \Lambda^0 $ & $ -\sqrt{\frac{3}{2}} c_1 V_{ud}^*$&
$\overline B^0\to   {F'}_{-1/2}^0  \overline \Sigma^0 $ & $ \frac{c_1 V_{ud}^*}{-\sqrt{2}}$\\
$\overline B^0\to   {F'}_{-1/2}^0  \overline n $ & $ c_3 V_{us}^*$&
$\overline B^0\to   F_{-3/2}^-  \overline \Sigma^+ $ & $ \left(c_1+c_3\right) V_{ud}^*$&
$\overline B^0\to   F_1^+  \overline \Sigma^- $ & $ c_2 V_{us}^*$\\
$\overline B^0\to   F_0^0  \overline \Lambda^0 $ & $ \frac{\left(c_2+c_3-2 c_4\right) V_{us}^*}{\sqrt{6}}$&
$\overline B^0\to   F_0^0  \overline \Sigma^0 $ & $ \frac{\left(c_2-c_3\right) V_{us}^*}{\sqrt{2}}$&
$\overline B^0\to   F_0^0  \overline \Xi^0 $ & $ \left(c_1+c_4\right) V_{ud}^*$\\
$\overline B^0\to   F_{-1}^-  \overline \Sigma^+ $ & $ c_3 V_{us}^*$&
$\overline B^0\to   F_{-1}^-  \overline \Xi^+ $ & $ c_1 V_{ud}^*$&
$\overline B^0_s\to   F_{-1/2}^+  \overline p $ & $ c_2 V_{ud}^*$\\
$\overline B^0_s\to   {F'}_0^+  \overline \Sigma^- $ & $ c_2 V_{ud}^*$&
$\overline B^0_s\to   {F'}_0^+  \overline p $ & $ -c_2 V_{us}^*$&
$\overline B^0_s\to   F_0^+  \overline \Sigma^- $ & $ c_4 V_{ud}^*$\\
$\overline B^0_s\to   F_0^+  \overline p $ & $ \left(c_1-c_2+c_4\right) V_{us}^*$&
$\overline B^0_s\to   F_{-1}^0  \overline \Lambda^0 $ & $ \frac{\left(c_2-2 c_3+c_4\right) V_{ud}^*}{\sqrt{6}}$&
$\overline B^0_s\to   F_{-1}^0  \overline \Sigma^0 $ & $ \frac{\left(c_2-c_4\right) V_{ud}^*}{\sqrt{2}}$\\
$\overline B^0_s\to   F_{-1}^0  \overline n $ & $ \left(c_1+c_4\right) V_{us}^*$&
$\overline B^0_s\to   F_{1/2}^+  \overline \Sigma^- $ & $ c_1 V_{us}^*$&
$\overline B^0_s\to   {F'}_{1/2}^+  \overline \Sigma^- $ & $ c_2 V_{us}^*$\\
$\overline B^0_s\to   F_{-1/2}^0  \overline \Sigma^0 $ & $ -\sqrt{2} c_1 V_{us}^*$&
$\overline B^0_s\to   F_{-1/2}^0  \overline \Xi^0 $ & $ c_3 V_{ud}^*$&
$\overline B^0_s\to   {F'}_{-1/2}^0  \overline \Lambda^0 $ & $ \frac{\left(c_2-3 c_1-2 \left(c_3+c_4\right)\right) V_{us}^*}{\sqrt{6}}$\\
$\overline B^0_s\to   {F'}_{-1/2}^0  \overline \Sigma^0 $ & $ \frac{\left(c_2-c_1\right) V_{us}^*}{\sqrt{2}}$&
$\overline B^0_s\to   {F'}_{-1/2}^0  \overline \Xi^0 $ & $ c_4 V_{ud}^*$&
$\overline B^0_s\to   F_{-3/2}^-  \overline \Sigma^+ $ & $ c_1 V_{us}^*$\\
$\overline B^0_s\to   F_{-3/2}^-  \overline \Xi^+ $ & $ c_3 V_{ud}^*$&
$\overline B^0_s\to   F_0^0  \overline \Xi^0 $ & $ \left(c_1+c_3\right) V_{us}^*$&
$\overline B^0_s\to   F_{-1}^-  \overline \Xi^+ $ & $ \left(c_1+c_3\right) V_{us}^*$\\\hline
$B^-\to   F_0^+  \overline \Delta^{--} $ & $ -\left(d_1+d_2\right) V_{ud}^*$&
$B^-\to   F_{-1}^0  \overline \Delta^{-} $ & $ \frac{\left(d_1+d_2\right) V_{ud}^*}{-\sqrt{3}}$&
$B^-\to   F_{1/2}^+  \overline \Delta^{--} $ & $ \left(d_1+d_2\right) V_{us}^*$\\
$B^-\to   F_{-1/2}^0  \overline \Delta^{-} $ & $ \frac{2 \left(d_1+d_2\right)V_{us}^*}{\sqrt{3}}$&
$B^-\to   F_{-1/2}^0  \overline \Sigma^{\prime-} $ & $ \frac{\left(d_1+d_2\right) V_{ud}^*}{-\sqrt{3}}$&
$B^-\to   {F'}_{-1/2}^0  \overline \Delta^{-} $ & $ \frac{\left(d_1+d_2\right) V_{us}^*}{\sqrt{3}}$\\
$B^-\to   {F'}_{-1/2}^0  \overline \Sigma^{\prime-} $ & $ \frac{2 \left(d_1+d_2\right) V_{ud}^*}{-\sqrt{3}}$&
$B^-\to   F_{-3/2}^-  \overline \Delta^{0} $ & $ \frac{\left(d_1+d_2\right) V_{us}^*}{-\sqrt{3}}$&
$B^-\to   F_{-3/2}^-  \overline \Sigma^{\prime0} $ & $ \frac{\left(d_1+d_2\right) V_{ud}^*}{\sqrt{6}}$\\
$B^-\to   F_0^0  \overline \Sigma^{\prime-} $ & $ \frac{\left(d_1+d_2\right) V_{us}^*}{\sqrt{3}}$&
$B^-\to   F_{-1}^-  \overline \Sigma^{\prime0} $ & $ \frac{\left(d_1+d_2\right) V_{us}^*}{-\sqrt{6}}$&
$B^-\to   F_{-1}^-  \overline \Xi^{\prime0} $ & $ \frac{\left(d_1+d_2\right) V_{ud}^*}{\sqrt{3}}$\\
$\overline B^0\to   F_1^{++}  \overline \Delta^{--} $ & $ -d_1 V_{ud}^*$&
$\overline B^0\to   {F'}_0^+  \overline \Delta^{-} $ & $ \frac{d_1 V_{ud}^*}{-\sqrt{3}}$&
$\overline B^0\to   F_0^+  \overline \Delta^{-} $ & $ \frac{d_2 V_{ud}^*}{-\sqrt{3}}$\\
$\overline B^0\to   F_{-1}^0  \overline \Delta^{0} $ & $ \frac{d_2 V_{ud}^*}{-\sqrt{3}}$&
$\overline B^0\to   F_{3/2}^{++}  \overline \Delta^{--} $ & $ d_1 V_{us}^*$&
$\overline B^0\to   F_{1/2}^+  \overline \Delta^{-} $ & $ \frac{\left(d_2-2 d_1\right) V_{us}^*}{\sqrt{3}}$\\
$\overline B^0\to   F_{1/2}^+  \overline \Sigma^{\prime-} $ & $ \frac{d_1 V_{ud}^*}{\sqrt{3}}$&
$\overline B^0\to   {F'}_{1/2}^+  \overline \Delta^{-} $ & $ \frac{d_1 V_{us}^*}{-\sqrt{3}}$&
$\overline B^0\to   {F'}_{1/2}^+  \overline \Sigma^{\prime-} $ & $ \frac{d_1 V_{ud}^*}{-\sqrt{3}}$\\
$\overline B^0\to   F_{-1/2}^0  \overline \Delta^{0} $ & $ \frac{\left(d_1-2 d_2\right) V_{us}^*}{-\sqrt{3}}$&
$\overline B^0\to   F_{-1/2}^0  \overline \Sigma^{\prime0} $ & $\frac{\left(d_1-d_2\right) V_{ud}^*}{\sqrt{6}}$&
$\overline B^0\to   {F'}_{-1/2}^0  \overline \Delta^{0} $ & $ \frac{d_2 V_{us}^*}{\sqrt{3}}$\\
$\overline B^0\to   {F'}_{-1/2}^0  \overline \Sigma^{\prime0} $ & $ -\sqrt{\frac{2}{3}} d_2 V_{ud}^*$&
$\overline B^0\to   F_{-3/2}^-  \overline \Delta^{+} $ & $ -d_2 V_{us}^*$&
$\overline B^0\to   F_{-3/2}^-  \overline \Sigma^{\prime+} $ & $ \frac{d_2 V_{ud}^*}{\sqrt{3}}$\\
$\overline B^0\to   F_1^+  \overline \Sigma^{\prime-} $ & $ \frac{d_1 V_{us}^*}{\sqrt{3}}$&
$\overline B^0\to   F_0^0  \overline \Sigma^{\prime0} $ & $ \frac{\left(d_2-d_1\right) V_{us}^*}{\sqrt{6}}$&
$\overline B^0\to   F_0^0  \overline \Xi^{\prime0} $ & $ \frac{d_1 V_{ud}^*}{\sqrt{3}}$\\
$\overline B^0\to   F_{-1}^-  \overline \Sigma^{\prime+} $ & $ \frac{d_2 V_{us}^*}{-\sqrt{3}}$&
$\overline B^0\to   F_{-1}^-  \overline \Xi^{\prime+} $ & $ \frac{d_2 V_{ud}^*}{\sqrt{3}}$&
$\overline B^0_s\to   F_{1/2}^{++}  \overline \Delta^{--} $ & $ -d_1 V_{ud}^*$\\
$\overline B^0_s\to   F_{-1/2}^+  \overline \Delta^{-} $ & $ \frac{d_1 V_{ud}^*}{-\sqrt{3}}$&
$\overline B^0_s\to   F_1^{++}  \overline \Delta^{--} $ & $ d_1 V_{us}^*$&
$\overline B^0_s\to   {F'}_0^+  \overline \Delta^{-} $ & $ \frac{d_1 V_{us}^*}{\sqrt{3}}$\\
$\overline B^0_s\to   {F'}_0^+  \overline \Sigma^{\prime-} $ & $ \frac{d_1 V_{ud}^*}{\sqrt{3}}$&
$\overline B^0_s\to   F_0^+  \overline \Delta^{-} $ & $ \frac{d_1 V_{us}^*}{-\sqrt{3}}$&
$\overline B^0_s\to   F_0^+  \overline \Sigma^{\prime-} $ & $ \frac{\left(2 d_1-d_2\right) V_{ud}^*}{\sqrt{3}}$\\
$\overline B^0_s\to   F_{-1}^0  \overline \Delta^{0} $ & $ \frac{d_1V_{us}^*}{-\sqrt{3}}$&
$\overline B^0_s\to   F_{-1}^0  \overline \Sigma^{\prime0} $ & $ \frac{\left(d_1-d_2\right) V_{ud}^*}{\sqrt{6}}$&
$\overline B^0_s\to   F_{1/2}^+  \overline \Sigma^{\prime-} $ & $ \frac{d_2 V_{us}^*}{\sqrt{3}}$\\
$\overline B^0_s\to   {F'}_{1/2}^+  \overline \Sigma^{\prime-} $ & $ \frac{d_1 V_{us}^*}{\sqrt{3}}$&
$\overline B^0_s\to   F_{-1/2}^0  \overline \Sigma^{\prime0} $ & $ \sqrt{\frac{2}{3}} d_2 V_{us}^*$&
$\overline B^0_s\to   F_{-1/2}^0  \overline \Xi^{\prime0} $ & $ \frac{d_2 V_{ud}^*}{-\sqrt{3}}$\\
$\overline B^0_s\to   {F'}_{-1/2}^0  \overline \Sigma^{\prime0} $ & $ \frac{\left(d_2-d_1\right) V_{us}^*}{\sqrt{6}}$&
$\overline B^0_s\to   {F'}_{-1/2}^0  \overline \Xi^{\prime0} $ & $ \frac{\left(d_1-2 d_2\right) V_{ud}^*}{\sqrt{3}}$&
$\overline B^0_s\to   F_{-3/2}^-  \overline \Sigma^{\prime+} $ & $ \frac{d_2 V_{us}^*}{-\sqrt{3}}$\\
$\overline B^0_s\to   F_{-3/2}^-  \overline \Xi^{\prime+} $ & $ \frac{d_2 V_{ud}^*}{\sqrt{3}}$&
$\overline B^0_s\to   F_0^0  \overline \Xi^{\prime0} $ & $ \frac{d_2 V_{us}^*}{\sqrt{3}}$&
$\overline B^0_s\to   F_{-1}^-  \overline \Xi^{\prime+} $ & $ \frac{d_2 V_{us}^*}{-\sqrt{3}}$\\
$\overline B^0_s\to   F_{-1}^-  \overline \Omega^+ $ & $ d_2 V_{ud}^*$&&&&\\\hline
\hline
\end{tabular}
\end{table}

\begin{table}
\caption{The productions of pentaquark $S_6$, light anti-baryons $({qqq})_{8/10}$ or anti-charm anti-baryon $(\bar{c}qq)_{3/\bar 6}$ from B mesons.}
\label{tab:P6full}\begin{tabular}{|cc|cc|cc|c|c}\hline\hline
channel & amplitude& channel & amplitude&channel & amplitude  \\\hline
$B^-\to   S_{-1}^0  \overline \Lambda_{\bar c}^- $ & $ -a_1 V_{cd}^*$&
$B^-\to   S_{-1/2}^0  \overline \Lambda_{\bar c}^- $ & $ -a_1 V_{cs}^*$&
$B^-\to   S_{-1/2}^0  \overline \Xi_{\bar c}^- $ & $ -a_1 V_{cd}^*$\\
$B^-\to   S_{0}^0  \overline \Xi_{\bar c}^- $ & $ -a_1 V_{cs}^*$&
$\overline B^0\to   S_{0}^+  \overline \Lambda_{\bar c}^- $ & $ a_1 V_{cd}^*$&
$\overline B^0\to   S_{1/2}^+  \overline \Lambda_{\bar c}^- $ & $ a_1 V_{cs}^*$\\
$\overline B^0\to   S_{-1/2}^0  \overline \Xi_{\bar c}^0 $ & $ -a_1 V_{cd}^*$&
$\overline B^0\to   S_{0}^0  \overline \Xi_{\bar c}^0 $ & $ -a_1 V_{cs}^*$&
$\overline B^0_s\to   S_{0}^+  \overline \Xi_{\bar c}^- $ & $ a_1 V_{cd}^*$\\
$\overline B^0_s\to   S_{1/2}^+  \overline \Xi_{\bar c}^- $ & $ a_1 V_{cs}^*$&
$\overline B^0_s\to   S_{-1}^0  \overline \Xi_{\bar c}^0 $ & $ a_1 V_{cd}^*$&
$\overline B^0_s\to   S_{-1/2}^0  \overline \Xi_{\bar c}^0 $ & $ a_1 V_{cs}^*$\\
\hline
$B^-\to   S_{0}^+  \overline \Sigma_{\bar c}^{--} $ & $ b_1 V_{cd}^*$&
$B^-\to   S_{1/2}^+  \overline \Sigma_{\bar c}^{--} $ & $ b_1 V_{cs}^*$&
$B^-\to   S_{-1}^0  \overline \Sigma_{\bar c}^{-} $ & $ \frac{b_1 V_{cd}^*}{\sqrt{2}}$\\
$B^-\to   S_{-1/2}^0  \overline \Sigma_{\bar c}^{-} $ & $ \frac{b_1 V_{cs}^*}{\sqrt{2}}$&$B^-\to   S_{-1/2}^0  \overline \Xi_{\bar c}^{\prime-} $ & $ \frac{b_1 V_{cd}^*}{\sqrt{2}}$&$B^-\to   S_{0}^0  \overline \Xi_{\bar c}^{\prime-} $ & $ \frac{b_1 V_{cs}^*}{\sqrt{2}}$\\
$\overline B^0\to   S_1^{++}  \overline \Sigma_{\bar c}^{--} $ & $ b_2 V_{cd}^*$&
$\overline B^0\to   S_{0}^+  \overline \Sigma_{\bar c}^{-} $ & $ \frac{\left(b_1+2 b_2\right) V_{cd}^*}{\sqrt{2}}$&
$\overline B^0\to   S_{1/2}^+  \overline \Sigma_{\bar c}^{-} $ & $ \frac{b_1 V_{cs}^*}{\sqrt{2}}$\\
$\overline B^0\to   S_{1/2}^+  \overline \Xi_{\bar c}^{\prime-} $ & $ \sqrt{2} b_2 V_{cd}^*$&
$\overline B^0\to   S_{-1}^0  \overline \Sigma_{\bar c}^{0} $ & $ \left(b_1+b_2\right) V_{cd}^*$&
$\overline B^0\to   S_{-1/2}^0  \overline \Sigma_{\bar c}^{0} $ & $ b_1 V_{cs}^*$\\
$\overline B^0\to   S_{-1/2}^0  \overline \Xi_{\bar c}^{\prime0} $ & $ \frac{\left(b_1+2 b_2\right) V_{cd}^*}{\sqrt{2}}$&
$\overline B^0\to   S_{0}^0  \overline \Xi_{\bar c}^{\prime0} $ & $ \frac{b_1 V_{cs}^*}{\sqrt{2}}$&
$\overline B^0\to   S_{0}^0  \overline \Omega_{\bar c}^{0} $ & $ b_2 V_{cd}^*$\\
$\overline B^0_s\to   S_1^{++}  \overline \Sigma_{\bar c}^{--} $ & $ b_2 V_{cs}^*$&
$\overline B^0_s\to   S_{0}^+  \overline \Sigma_{\bar c}^{-} $ & $ \sqrt{2} b_2 V_{cs}^*$&
$\overline B^0_s\to   S_{0}^+  \overline \Xi_{\bar c}^{\prime-} $ & $ \frac{b_1 V_{cd}^*}{\sqrt{2}}$\\
$\overline B^0_s\to   S_{1/2}^+  \overline \Xi_{\bar c}^{\prime-} $ & $ \frac{\left(b_1+2 b_2\right) V_{cs}^*}{\sqrt{2}}$&$\overline B^0_s\to   S_{-1}^0  \overline \Sigma_{\bar c}^{0} $ & $ b_2 V_{cs}^*$&$\overline B^0_s\to   S_{-1}^0  \overline \Xi_{\bar c}^{\prime0} $ & $ \frac{b_1 V_{cd}^*}{\sqrt{2}}$\\
$\overline B^0_s\to   S_{-1/2}^0  \overline \Xi_{\bar c}^{\prime0} $ & $ \frac{\left(b_1+2 b_2\right) V_{cs}^*}{\sqrt{2}}$&$\overline B^0_s\to   S_{-1/2}^0  \overline \Omega_{\bar c}^{0} $ & $ b_1 V_{cd}^*$&$\overline B^0_s\to   S_{0}^0  \overline \Omega_{\bar c}^{0} $ & $ \left(b_1+b_2\right) V_{cs}^*$\\\hline
$B^-\to   S_{-1}^0  \overline p $ & $ \left(c_2+c_3\right) V_{ud}^*$&
$B^-\to   S_{-1/2}^0  \overline \Sigma^- $ & $ -\left(c_2+c_3\right) V_{ud}^*$&
$B^-\to   S_{-1/2}^0  \overline p $ & $ \left(c_2+c_3\right) V_{us}^*$\\
$B^-\to   S_{0}^0  \overline \Sigma^- $ & $ -\left(c_2+c_3\right) V_{us}^*$&
$\overline B^0\to   S_{0}^+  \overline p $ & $ \left(c_1-c_3-c_4\right) V_{ud}^*$&
$\overline B^0\to   S_{1/2}^+  \overline \Sigma^- $ & $ -c_1 V_{ud}^*$\\
$\overline B^0\to   S_{1/2}^+  \overline p $ & $ -\left(c_3+c_4\right) V_{us}^*$&
$\overline B^0\to   S_{-1}^0  \overline n $ & $ \left(c_1+c_2-c_4\right) V_{ud}^*$&
$\overline B^0\to   S_{-1/2}^0  \overline \Lambda^0 $ & $ \frac{\left(2 c_4-3 c_1-c_2+c_3\right) V_{ud}^*}{\sqrt{6}}$\\
$\overline B^0\to   S_{-1/2}^0  \overline \Sigma^0 $ & $ \frac{\left(c_1+c_2+c_3\right) V_{ud}^*}{\sqrt{2}}$&
$\overline B^0\to   S_{-1/2}^0  \overline n $ & $ \left(c_2-c_4\right) V_{us}^*$&
$\overline B^0\to   S_{0}^0  \overline \Lambda^0 $ & $ \frac{\left(2 c_4-c_2+c_3\right) V_{us}^*}{\sqrt{6}}$\\
$\overline B^0\to   S_{0}^0  \overline \Sigma^0 $ & $ \frac{\left(c_2+c_3\right) V_{us}^*}{\sqrt{2}}$&
$\overline B^0\to   S_{0}^0  \overline \Xi^0 $ & $ -c_1V_{ud}^*$&
$\overline B^0_s\to   S_{0}^+  \overline \Sigma^- $ & $ \left(c_3+c_4\right) V_{ud}^*$\\
$\overline B^0_s\to   S_{0}^+  \overline p $ & $ c_1 V_{us}^*$&
$\overline B^0_s\to   S_{1/2}^+  \overline \Sigma^- $ & $ \left(c_4-c_1+c_3\right) V_{us}^*$&
$\overline B^0_s\to   S_{-1}^0  \overline \Lambda^0 $ & $ \frac{\left(c_4-2 c_2-c_3\right) V_{ud}^*}{\sqrt{6}}$\\
$\overline B^0_s\to   S_{-1}^0  \overline \Sigma^0 $ & $ \frac{\left(c_3+c_4\right) V_{ud}^*}{-\sqrt{2}}$&
$\overline B^0_s\to   S_{-1}^0  \overline n $ & $ c_1 V_{us}^*$&
$\overline B^0_s\to   S_{-1/2}^0  \overline \Lambda^0 $ & $ \frac{\left(3 c_1+2 c_2+c_3-c_4\right) V_{us}^*}{-\sqrt{6}}$\\
$\overline B^0_s\to   S_{-1/2}^0  \overline \Sigma^0 $ & $ \frac{\left(c_1-c_3-c_4\right) V_{us}^*}{\sqrt{2}}$&
$\overline B^0_s\to   S_{-1/2}^0  \overline \Xi^0 $ & $ \left(c_4-c_2\right) V_{ud}^*$&
$\overline B^0_s\to   S_{0}^0  \overline \Xi^0 $ & $ \left(c_4-c_1-c_2\right) V_{us}^*$\\\hline
\end{tabular}
\end{table}
The possible Hamiltonian for the production of pentaquark ground $6$ states from one $B$ meson, induced by the transition $b\to c \bar{c}d/s$ and $b\to c \bar{u}d/s$ in the quark level, can be writen directly as,
\begin{eqnarray}
 {\mathcal{H}}_{6} &=& a_{1}{B}_{i}(H_{3})^{j}({S}_{6})_{\{jk\}}(F_{c3})^{[ik]}+b_{1}{B}_{i}(H_{3})^{j}({S}_{6})_{\{jk\}}(F_{c\bar 6})^{\{ik\}}+b_{2}{B}_{i}(H_{3})^{i}{S}_{6})_{\{jk\}}(F_{c\bar 6})^{\{jk\}}\nonumber\\
 &+&c_{1}{B}_{i}(H_{8})_{j}^{i}(S_{6})_{\{kl\}}(F_{8})_{\alpha}^{k}\ \varepsilon^{\alpha jl}+ c_{2}{B}_{i}(H_{8})_{j}^{k}(S_{6})_{\{kl\}}(F_{8})_{\alpha}^{i}\ \varepsilon^{\alpha jl}+c_{3}{B}_{i}(H_{8})_{j}^{k}(S_{6})_{\{kl\}}(F_{8})_{\alpha}^{j}\ \varepsilon^{\alpha il}\nonumber\\&+&c_{4}{B}_{i}(H_{8})_{j}^{k}(S_{6})_{\{kl\}}(F_{8})_{\alpha}^{l}\ \varepsilon^{\alpha ij}.\label{eq:hamiltonian6}
 \end{eqnarray}
The production channels directly collected into Tab.~\ref{tab:P6full}, in addition, we show the fully production processes about all 15 states in the Tab.~\ref{tab:P15full1} and Tab.~\ref{tab:P15full2}.
The relations between different production widths can then be deduced, the complete results for sextet given as,
\begin{align*}
&
\Gamma(B^-\to S_{-1}^0 \overline \Lambda_{\bar c}^-)
= \Gamma(B^-\to S_{-1/2}^0 \overline \Xi_{\bar c}^-)
= \Gamma(\overline B^0\to S_{-1/2}^0 \overline \Xi_{\bar c}^0)= \Gamma(\overline B^0_s\to S_{-1}^0 \overline \Xi_{\bar c}^0)\\
&
=\Gamma(\overline B^0\to S_{0}^+ \overline \Lambda_{\bar c}^-)
=\Gamma(\overline B^0_s\to S_{0}^+ \overline \Xi_{\bar c}^-),\\
&
\Gamma(B^-\to S_{-1/2}^0 \overline \Lambda_{\bar c}^-)
=\Gamma(B^-\to S_{0}^0 \overline \Xi_{\bar c}^-) =\Gamma(\overline B^0\to S_{0}^0 \overline \Xi_{\bar c}^0)
=\Gamma(\overline B^0\to S_{1/2}^+ \overline \Lambda_{\bar c}^-)\\
&
= \Gamma(\overline B^0_s\to S_{1/2}^+ \overline \Xi_{\bar c}^-)
= \Gamma(\overline B^0_s\to S_{-1/2}^0 \overline \Xi_{\bar c}^0),\\
&\Gamma(B^-\to S_{0}^+ \overline \Sigma_{\bar c}^{--})
= 2\Gamma(B^-\to S_{-1}^0 \overline \Sigma_{\bar c}^{-})
=2\Gamma(B^-\to S_{-1/2}^0 \overline \Xi_{\bar c}^{\prime-})
=2\Gamma(B^-\to S_{0}^0 \overline \Xi_{\bar c}^{\prime-})\\
&
=\Gamma(\overline B^0\to S_{-1/2}^0 \overline \Sigma_{\bar c}^{0})=2\Gamma(\overline B^0\to S_{1/2}^+ \overline \Sigma_{\bar c}^{-})
=2\Gamma(\overline B^0\to S_{0}^0 \overline \Xi_{\bar c}^{\prime0}
=2\Gamma(\overline B^0_s\to S_{0}^+ \overline \Xi_{\bar c}^{\prime-})\\
&
=2\Gamma(\overline B^0_s\to S_{-1}^0 \overline \Xi_{\bar c}^{\prime0})
=\Gamma(\overline B^0_s\to S_{-1/2}^0 \overline \Omega_{\bar c}^{0})=\Gamma(B^-\to S_{1/2}^+ \overline \Sigma_{\bar c}^{--})
= 2\Gamma(\overline B^0\to S_{1/2}^+ \overline \Sigma_{\bar c}^{-}),\\
&\Gamma(\overline B^0\to S_{0}^+ \overline \Sigma_{\bar c}^{-})
= \Gamma(\overline B^0\to S_{-1/2}^0 \overline \Xi_{\bar c}^{\prime0}),\quad
\Gamma(\overline B^0\to S_{0}^+ \overline \Sigma_{\bar c}^{-})
= \Gamma(\overline B^0\to S_{-1/2}^0 \overline \Xi_{\bar c}^{\prime0}),\\
&
\Gamma(\overline B^0_s\to S_1^{++} \overline \Sigma_{\bar c}^{--})= \frac{1}{2}\Gamma(\overline B^0_s\to S_{0}^+ \overline \Sigma_{\bar c}^{-})
=\Gamma(\overline B^0_s\to S_{-1}^0 \overline \Sigma_{\bar c}^{0}),\\
&2\Gamma(\overline B^0\to S_1^{++} \overline \Sigma_{\bar c}^{--})
= \Gamma(\overline B^0\to S_{1/2}^+ \overline \Xi_{\bar c}^{\prime-})
=2\Gamma(\overline B^0\to S_{0}^0 \overline \Omega_{\bar c}^{0}),\\
&
\Gamma(B^-\to S_{-1/2}^0 \overline p)
= \Gamma(B^-\to S_{0}^0 \overline \Sigma^-)
= 2\Gamma(\overline B^0\to S_{0}^0 \overline \Sigma^0)
=\Gamma(B^-\to S_{0}^0 \overline \Sigma^-),\\
&
\Gamma(B^-\to S_{-1}^0 \overline p)= \Gamma(B^-\to S_{-1/2}^0 \overline \Sigma^-), \quad \Gamma(\overline B^0\to S_{1/2}^+ \overline \Sigma^-)= \Gamma(\overline B^0\to S_{0}^0 \overline \Xi^0),\\
&
 \Gamma(\overline B^0_s\to S_{0}^+ \overline \Sigma^-)= 2\Gamma(\overline B^0_s\to S_{-1}^0 \overline \Sigma^0), \quad \Gamma(\overline B^0_s\to S_{0}^+ \overline p)= \Gamma(\overline B^0_s\to S_{-1}^0 \overline n),\\
&
\Gamma(\overline B^0_s\to S_{1/2}^+ \overline \Xi_{\bar c}^{\prime-})
=\Gamma(\overline B^0_s\to S_{-1/2}^0 \overline \Xi_{\bar c}^{\prime0}),
\Gamma(\overline B^0_s\to S_{1/2}^+ \overline \Sigma^-)= 2\Gamma(\overline B^0_s\to S_{-1/2}^0 \overline \Sigma^0).
\end{align*}
The production width relations of 15 states are gathered as,
\begin{align*}
&\Gamma(B^-\to F_{-1}^0 \bar \Lambda_{\bar{c}}^-)
=\Gamma(\overline B^0\to F_{1/2}^+ \bar \Xi_{\bar{c}}^-)
= \Gamma(\overline B^0_s\to F_{-1}^0 \overline \Xi_{\bar{c}}^0)
= \Gamma(\overline B^0\to {F'}_0^+ \bar \Lambda_{\bar{c}}^-)\\
&=\Gamma(\overline B^0\to {F'}_{1/2}^+ \bar \Xi_{\bar{c}}^-)
=\Gamma(\overline B^0\to F_{-1/2}^0 \overline \Xi_{\bar{c}}^0)
=\Gamma(\overline B^0_s\to {F'}_0^+ \bar \Xi_{\bar{c}}^-)
=\Gamma(\overline B^0_s\to F_{-1/2}^+ \bar \Lambda_{\bar{c}}^-)\\
&=\Gamma(B^-\to F_{-3/2}^- \overline \Xi_{\bar{c}}^0)
= \Gamma(B^-\to F_{-1/2}^0 \bar \Xi_{\bar{c}}^-)
,\\
 &\Gamma(B^-\to {F'}_{-1/2}^0 \bar \Lambda_{\bar{c}}^-)
=\Gamma(B^-\to F_0^0 \bar \Xi_{\bar{c}}^-)
=\Gamma(B^-\to F_{-1}^- \overline \Xi_{\bar{c}}^0)
=\Gamma(\overline B^0\to F_0^0 \overline \Xi_{\bar{c}}^0)\\
&=\Gamma(\overline B^0\to {F'}_{1/2}^+ \bar \Lambda_{\bar{c}}^-)
=\Gamma(\overline B^0\to F_1^+ \bar \Xi_{\bar{c}}^-)
=\Gamma(\overline B^0_s\to {F'}_{1/2}^+ \bar \Xi_{\bar{c}}^-)
=\Gamma(\overline B^0_s\to {F'}_0^+ \bar \Lambda_{\bar{c}}^-)\\
&=\Gamma(\overline B^0_s\to F_0^+ \bar \Lambda_{\bar{c}}^-)
 = \Gamma(\overline B^0_s\to {F'}_{-1/2}^0 \overline \Xi_{\bar{c}}^0)
,\\
&\Gamma(B^-\to   F_0^+  \bar \Sigma_{\bar{c}}^{--})
=2\Gamma(B^-\to   F_{-1}^0  \bar \Sigma_{\bar{c}}^{-})
=2\Gamma(B^-\to   F_{-1/2}^0  \bar \Xi_{\bar{c}}^{\prime-})
=\frac{1}{2}\Gamma(B^-\to   {F'}_{-1/2}^0  \bar \Xi_{\bar{c}}^{\prime-})\\
&=2\Gamma(B^-\to   F_{-3/2}^-  \overline \Xi_{\bar{c}}^{\prime0})
=\Gamma(B^-\to   F_{-1}^-  \overline \Omega_{\bar{c}}^{0})
=\Gamma(\overline B^0\to   F_1^{++}  \bar \Sigma_{\bar{c}}^{--})
=2\Gamma(\overline B^0\to   {F'}_0^+  \bar \Sigma_{\bar{c}}^{-})\\
&=2\Gamma(\overline B^0\to   F_{1/2}^+  \bar \Xi_{\bar{c}}^{\prime-})
=2\Gamma(\overline B^0\to   {F'}_{1/2}^+  \bar \Xi_{\bar{c}}^{\prime-})
=2\Gamma(\overline B^0\to   F_{-1/2}^0  \overline \Xi_{\bar{c}}^{\prime0})
=\Gamma(\overline B^0\to   F_0^0  \overline \Omega_{\bar{c}}^{0})\\
&=\Gamma(\overline B^0_s\to   F_{1/2}^{++}  \bar \Sigma_{\bar{c}}^{--})
=2\Gamma(\overline B^0_s\to   F_{-1/2}^+  \bar \Sigma_{\bar{c}}^{-})
=2\Gamma(\overline B^0_s\to   {F'}_0^+  \bar \Xi_{\bar{c}}^{\prime-})
=\frac{1}{2}\Gamma(\overline B^0_s\to   F_0^+  \bar \Xi_{\bar{c}}^{\prime-})\\
&=2\Gamma(\overline B^0_s\to   F_{-1}^0  \overline \Xi_{\bar{c}}^{\prime0})
=\Gamma(\overline B^0_s\to {F'}_{-1/2}^0 \overline \Omega_{\bar{c}}^{0}),\\
&\Gamma(B^-\to   F_{1/2}^+  \bar \Sigma_{\bar{c}}^{--})
=\frac{1}{2}\Gamma(B^-\to   F_{-1/2}^0  \bar \Sigma_{\bar{c}}^{-})=
2\Gamma(B^-\to   {F'}_{-1/2}^0  \bar \Sigma_{\bar{c}}^{-})
=\Gamma(B^-\to   F_{-3/2}^-  \overline \Sigma_{\bar{c}}^{0})\\
&=2\Gamma(B^-\to   F_0^0  \bar \Xi_{\bar{c}}^{\prime-})
=2\Gamma(B^-\to   F_{-1}^-  \overline \Xi_{\bar{c}}^{\prime0})
=\Gamma(\overline B^0\to   F_{3/2}^{++}  \bar \Sigma_{\bar{c}}^{--})
=\frac{1}{2}\Gamma(\overline B^0\to   F_{1/2}^+  \bar \Sigma_{\bar{c}}^{-})\\
&=2\Gamma(\overline B^0\to   {F'}_{1/2}^+  \bar \Sigma_{\bar{c}}^{-})
=\Gamma(\overline B^0\to   F_{-1/2}^0  \overline \Sigma_{\bar{c}}^{0})
=2\Gamma(\overline B^0\to   F_1^+  \bar \Xi_{\bar{c}}^{\prime-})
=2\Gamma(\overline B^0\to   F_0^0  \overline \Xi_{\bar{c}}^{\prime0})\\
&=\Gamma(\overline B^0_s\to   F_1^{++}  \bar \Sigma_{\bar{c}}^{--})
=2\Gamma(\overline B^0_s\to   {F'}_0^+  \bar \Sigma_{\bar{c}}^{-})
=2\Gamma(\overline B^0_s\to   F_0^+  \bar \Sigma_{\bar{c}}^{-})
=\Gamma(\overline B^0_s\to   F_{-1}^0  \overline \Sigma_{\bar{c}}^{0})\\
&=2\Gamma(\overline B^0_s\to   {F'}_{1/2}^+  \bar \Xi_{\bar{c}}^{\prime-})
=2\Gamma(\overline B^0_s\to   {F'}_{-1/2}^0  \overline \Xi_{\bar{c}}^{\prime0}),\\
&\Gamma(B^-\to F_{-1/2}^0 \overline p)
=\Gamma(B^-\to F_{-3/2}^- \overline n)
=\Gamma(\overline B^0\to F_{-1/2}^0 \overline n)
=\Gamma(\overline B^0\to F_{1/2}^+ \overline p),\\
&\Gamma(B^-\to {F'}_{-1/2}^0 \overline \Sigma^-)
=\Gamma(B^-\to F_{-1}^- \overline \Xi^0)
=\Gamma(\overline B^0_s\to F_0^+ \overline \Sigma^-)
=\Gamma(\overline B^0_s\to {F'}_{-1/2}^0 \overline \Xi^0),\\
&\Gamma(\overline B^0\to {F'}_0^+ \overline p)=
\Gamma(\overline B^0\to {F'}_{1/2}^+ \overline \Sigma^-)
= \Gamma(\overline B^0_s\to {F'}_0^+ \overline \Sigma^-)
=\Gamma(\overline B^0_s\to F_{-1/2}^+ \overline p),\\
&\Gamma(\overline B^0\to F_0^+ \overline p)=
\frac{2}{3}\Gamma(\overline B^0\to {F'}_{-1/2}^0 \overline \Lambda^0)
=2\Gamma(\overline B^0\to {F'}_{-1/2}^0 \overline \Sigma^0)
=\Gamma(\overline B^0\to F_{-1}^- \overline \Xi^+),\\
&\Gamma(\overline B^0\to {F'}_{1/2}^+ \overline p)=
\Gamma(\overline B^0\to F_1^+ \overline \Sigma^-)=
\Gamma(\overline B^0_s\to {F'}_{1/2}^+ \overline \Sigma^-)
=\Gamma(\overline B^0_s\to {F'}_0^+ \overline p),\\
&\Gamma(\overline B^0_s\to F_{1/2}^+ \overline \Sigma^-)
= \Gamma(\overline B^0_s\to F_{-3/2}^- \overline \Sigma^+)
= \frac{1}{2}\Gamma(\overline B^0_s\to F_{-1/2}^0 \overline \Sigma^0),\\
&\Gamma(B^-\to   F_0^+  \overline \Delta^{--})
=3\Gamma(B^-\to   F_{-1}^0  \overline \Delta^{-})
=\frac{3}{4}\Gamma(B^-\to   F_{-1/2}^0  \overline \Sigma^{\prime-})
=3\Gamma(B^-\to   {F'}_{-1/2}^0  \overline \Sigma^{\prime-})\\
&=6\Gamma(B^-\to   F_{-3/2}^-  \overline \Sigma^{\prime0})
=3\Gamma(B^-\to   F_{-1}^-  \overline \Xi^{\prime0}),\\
&\Gamma(B^-\to   F_{1/2}^+  \overline \Delta^{--})
=\frac{3}{4}\Gamma(B^-\to   F_{-1/2}^0  \overline \Delta^{-})
=3\Gamma(B^-\to   {F'}_{-1/2}^0  \overline \Delta^{-})
=3\Gamma(B^-\to   F_{-3/2}^-  \overline \Delta^{0})\\
&=3\Gamma(B^-\to   F_0^0  \overline \Sigma^{\prime-})
=6\Gamma(B^-\to   F_{-1}^-  \overline \Sigma^{\prime0}),\\
&\Gamma(\overline B^0\to   F_1^{++}  \overline \Delta^{--})
=3\Gamma(\overline B^0\to   {F'}_0^+  \overline \Delta^{-})
=3\Gamma(\overline B^0\to   F_{1/2}^+  \overline \Sigma^{\prime-})
=4\Gamma(\overline B^0\to   {F'}_{1/2}^+  \overline \Sigma^{\prime-})\\
&=3\Gamma(\overline B^0\to   F_0^0  \overline \Xi^{\prime0})
=\Gamma(\overline B^0_s\to   F_{1/2}^{++}  \overline \Delta^{--} )
=3\Gamma(\overline B^0_s\to   F_{-1/2}^+  \overline \Delta^{-})
=3\Gamma(\overline B^0_s\to   {F'}_0^+  \overline \Sigma^{\prime-}),\\
&\Gamma(\overline B^0\to   F_0^+  \overline \Delta^{-})
=\Gamma(\overline B^0\to   F_{-1}^0  \overline \Delta^{0})
=\frac{1}{2}\Gamma(\overline B^0\to   {F'}_{-1/2}^0  \overline \Sigma^{\prime0})
=\Gamma(\overline B^0\to   F_{-3/2}^-  \overline \Sigma^{\prime+})\\
&=\Gamma(\overline B^0\to   F_{-1}^-  \overline \Xi^{\prime+})
=\Gamma(\overline B^0_s\to   F_{-1/2}^0  \overline \Xi^{\prime0} )
=\Gamma(\overline B^0_s\to   F_{-3/2}^-  \overline \Xi^{\prime+})
=\frac{1}{3}\Gamma(\overline B^0_s\to   F_{-1}^-  \overline \Omega^+),\\
&\Gamma(\overline B^0\to   F_{3/2}^{++}  \overline \Delta^{--})
=3\Gamma(\overline B^0\to   {F'}_{1/2}^+  \overline \Delta^{-})
=3\Gamma(\overline B^0\to   F_1^+  \overline \Sigma^{\prime-})
=\Gamma(\overline B^0_s\to   F_1^{++}  \overline \Delta^{--})\\
&=3\Gamma(\overline B^0_s\to   {F'}_0^+  \overline \Delta^{-} )
=3\Gamma(\overline B^0_s\to   F_0^+  \overline \Delta^{-})
=3\Gamma(\overline B^0_s\to   F_{-1}^0  \overline \Delta^{0})
=3\Gamma(\overline B^0_s\to   {F'}_{1/2}^+  \overline \Sigma^{\prime-}),\\
&\Gamma(\overline B^0\to   {F'}_{-1/2}^0  \overline \Delta^{0})
=\frac{1}{3}\Gamma(\overline B^0\to   F_{-3/2}^-  \overline \Delta^{+})
=\Gamma(\overline B^0\to   F_{-1}^-  \overline \Sigma^{\prime+})
=\Gamma(\overline B^0_s\to   F_{1/2}^+  \overline \Sigma^{\prime-})\\
&=\frac{1}{2}\Gamma(\overline B^0_s\to   F_{-1/2}^0  \overline \Sigma^{\prime0})
=\Gamma(\overline B^0_s\to   F_{-3/2}^-  \overline \Sigma^{\prime+})
=\Gamma(\overline B^0_s\to   F_0^0  \overline \Xi^{\prime0})
=\Gamma(\overline B^0_s\to   F_{-1}^-  \overline \Xi^{\prime+}),\\
&\Gamma(\overline B^0_s\to   F_{-1}^0  \overline \Sigma^{\prime0})
=\Gamma(\overline B^0\to   F_{-1/2}^0  \overline \Sigma^{\prime0}),\Gamma(\overline B^0_s\to   {F'}_{-1/2}^0  \overline \Sigma^{\prime0})
=\Gamma(\overline B^0\to   F_0^0  \overline \Sigma^{\prime0}),\\
&\Gamma(\overline B^0\to F_{-1}^0 \overline n)=
\Gamma(\overline B^0\to F_{-3/2}^- \overline \Sigma^+),\Gamma(\overline B^0\to {F'}_{-1/2}^0 \overline n)=
\Gamma(\overline B^0\to F_{-1}^- \overline \Sigma^+),\\
&\Gamma(\overline B^0_s\to F_{-1/2}^0 \overline \Xi^0)=
\Gamma(\overline B^0_s\to F_{-3/2}^- \overline \Xi^+),\Gamma(\overline B^0_s\to F_0^0 \overline \Xi^0)=
\Gamma(\overline B^0_s\to F_{-1}^- \overline \Xi^+),\\
&\Gamma(B^-\to F_{-1/2}^0 \overline \Sigma^-)
= 2\Gamma(B^-\to F_{-3/2}^- \overline \Sigma^0),\Gamma(B^-\to F_0^0 \overline \Sigma^-)=
 2\Gamma(B^-\to F_{-1}^- \overline \Sigma^0).
\end{align*}
\section*{Acknowledgments}
This work is supported in part by the Fundamental Funds for Key disciplines in Physics with No.2022WLXK05, 2022WLXK15 and the National Natural Science Foundation of China under Grant No. 12005294.

\end{document}